\documentclass[preprint,aps,amsmath,superscriptaddress,nofootinbib,tightenlines,showpacs]{revtex4}
\usepackage{amsmath}
\usepackage{amsfonts}
\usepackage{amssymb}
\usepackage{latexsym}
\usepackage{graphicx}
\usepackage{bm}
\usepackage{epsfig}
\usepackage[english]{babel}

\def\sign{\mbox{sign}}
\begin{document}


\title{1/R Correction to Gravity in the Early Universe}
\author{Shi Pi\footnote{Electronic address: spi@pku.edu.cn}}
\affiliation{Department of Physics and
State Key Laboratory of Nuclear Physics and Technology, Peking University,\\
Beijing 100871, China}
\author{Tower Wang\footnote{Electronic address: wangtao218@pku.edu.cn}}
\affiliation{Center for High-Energy Physics, Peking University,\\
Beijing 100871, China\\ \vspace{0.2cm}}
\date{\today\\ \vspace{1cm}}
\begin{abstract}
To explain the accelerated expansion of the late universe, the $1/R$
correction to Einstein gravity is usually considered, where $R$ is
the Ricci scalar. This correction term, if stable, is generally
believed to be negligible during inflation. However, if the $1/R$
term is inflaton-dependent, it will dramatically change the story of
inflation. The entropy perturbation will naturally appear and drive
the evolution of curvature perturbation outside the Hubble horizon.
In a large class of models, the entropy perturbation can be made
nearly scale-invariant. In Einstein gravity the single-field
inflation with a quartic potential has been ruled out by recent
observations, but it revives when the $1/R$ term is turned on. The
evolution of non-Gaussianities on large scale are also studied and
applied to inflation with $1/R$ correction. In some specific models,
a large non-Gaussianity can be naturally generated outside the
horizon. Recent study ruled out almost all $f(R)$ models during
matter dominated phase. Taking this into consideration, we are left
with a limited class of model which recovers the Einstein gravity
soon after reheating.

\end{abstract}


\maketitle

\newpage



\section{Introduction}\label{sect-intro}
Despite great achievements of Einstein's gravity theory, numerous
versions of its modification or extension have been proposed in the
last and this century. Some proposals came and went, while others
were tightly constrained by observations \cite{Turyshev:2008ur}. One
of the modern motivations for modifying Einstein gravity is
attempting to explain the accelerated expansion of the late universe
\cite{Riess:1998cb,Perlmutter:1998np,Komatsu:2008hk}. Rather than
introducing a cosmological constant or an unknown dark energy, one
can explain the cosmic acceleration by designing a modified theory
of gravity, see typical models in
\cite{Dvali:2000hr,Carroll:2003wy,Li:2004rb} for instance. Among the
nonlinear modifications \cite{Nojiri:2006ri,Sotiriou:2008rp}, namely
$f(R)$ gravity theories, the most disputed one is a model with $1/R$
correction to the Einstein action \cite{Carroll:2003wy},
\begin{equation}\label{action-Rin}
S=\frac{M_p^2}{2}\int d^4x\sqrt{-g}\left(R-\frac{\mu^4}{R}\right)+\int d^4x\sqrt{-g}\mathcal{L}_M,
\end{equation}
though such a term looks bizarre from the viewpoint of effective
field theory.\footnote{Throughout this paper, we will mainly
following the conventions and notations of \cite{Ji:2009yw}. Some
conventions are gathered in the next section.}

In reference \cite{Carroll:2003wy}, it was assumed that the $1/R$
term was negligibly small in the early universe but gradually
reveals itself as the universe becomes more and more flat (the Ricci
scalar $R$ gets smaller and smaller) at the late time. However, in
general the parameter $\mu$ may depend on some matter fields and
therefore evolve along with the matter fields. Indeed, the general
coupling between gravity and the matter sector was considered in
recent investigations
\cite{Amendola:2006kh,Amendola:2006eh,Song:2006ej,Sawicki:2007tf,Nesseris:2008mq,Sadeghi:2009qr,Bertolami:2009cd,Ito:2009nk}.
Especially, the parameter $\mu$ can be a function of the inflaton
field, which will induce a correction term to single-field
inflation,
\begin{equation}\label{action-gRin}
S=\int d^4x\sqrt{-g}\left[\frac{1}{2}M_p^2R+\frac{g(\varphi)}{2R}-\frac{1}{2}g^{\alpha\beta}\partial_{\alpha}\varphi\partial_{\beta}\varphi-V(\varphi)\right].
\end{equation}
We expect the correction term was not negligible during inflation
but decayed soon after inflation (during reheating). By
fine-tuning the coupling $g(\varphi)$, one may also expect action
(\ref{action-gRin}) reproduces (\ref{action-Rin}) in the late
universe. Please refer to \cite{Liddle:2008bm} for a delicate
model unifying inflaton, dark matter and dark energy with a single
field.

In fact, the action (\ref{action-gRin}) is just a special case of
the $f(\varphi,R)$ generalized gravity, see
\cite{Ji:2009yw,Hwang:1990re,Hwang:1996np,Hwang:1996bc,Hwang:1997uc,Hwang:2005hb,Chen:2006wn}
and references therein. So we can employ the formalism recently
developed in \cite{Ji:2009yw} to deal with this model. In section
\ref{sect-general} we will collect the main results of
\cite{Ji:2009yw}, in a way as general as possible. The
non-Gaussianity in $f(\varphi,R)$ theory has not been studied
previously and deserves a separate investigation. But a
semi-quantitative analysis of this problem will be presented in
section \ref{sect-Gauss}. Then these formulas will be applied to
model (\ref{action-gRin}) in section \ref{sect-Rinverse}, where we
also find out the conditions for generating nearly scale-invariant
power spectra. Based on these results, we will in sections
\ref{sect-quadratic} and \ref{sect-quartic} study models with
specific potentials, \emph{i.e.},
$V(\varphi)=\frac{1}{2}m^2\varphi^2$ and
$V(\varphi)=\lambda\varphi^4$ respectively. The five-year Wilkinson
Microwave Anisotropy Probe (WMAP5) has ruled out single-field
inflation with $V(\varphi)=\lambda\varphi^4$ in Einstein gravity,
because in this model the tensor-to-scalar ratio $r_{T}$ is too
large and the the power spectrum of curvature perturbation
$P_{\mathcal{R}}$ is over-tilted
\cite{Komatsu:2008hk}.\footnote{See, however, reference
\cite{Ramirez:2009zs} for a counter example.} Interestingly, our
results will show that, in $V(\varphi)=\lambda\varphi^4$ model,
$r_{T}$ can be depressed by an order of magnitude while
$P_{\mathcal{R}}$ can be less tiled due to the $1/R$ term, hence the
model can pass the WMAP5 test. In the past few years, by considering
the coupling to matter in high redshift
\cite{Amendola:2006kh,Amendola:2006eh}, it was found that there are
instabilities in some branches of $f(R)$ gravity models
\cite{Song:2006ej,Sawicki:2007tf}. Therefore, in section
\ref{sect-stab}, we analyze the stability problem for our models and
its implication to post-inflation evolutions. We will conclude in
the last section after a few remarks on the possible loopholes and
the resulting uncertainty of our calculations. In appendix
\ref{app-3pt} we will derive some useful formulas for three-point
correlation functions of entropy and curvature perturbations. The
formulas developed in section \ref{sect-Gauss} and appendix
\ref{app-3pt} are very general and can be applied to other inflation
models with weakly coupled multiple fields.

\section{Inflation in Generalized Gravity}\label{sect-general}
There has been a lot of investigations on perturbation theory in
generalized gravity, where the Ricci scalar and a scalar field are
non-minimally coupled via an arbitrary function
\cite{Hwang:1990re,Hwang:1996np,Hwang:1996bc,Hwang:1997uc,Hwang:2005hb,Chen:2006wn}:
\begin{equation}\label{action-gg}
S=\int d^4x\sqrt{-g}\left[\frac{1}{2}f(\varphi,R)-\frac{1}{2}g^{\alpha\beta}\partial_{\alpha}\varphi\partial_{\beta}\varphi-V(\varphi)\right].
\end{equation}
But most of them are restricted to special cases with only one
degree of freedom, although it was believed that there should be two
degrees of freedom in general
\cite{Teyssandier:1983zz,Maeda:1988ab,Wands:1993uu}. In a recent
research \cite{Ji:2009yw}, such an $f(\varphi,R)$ theory was
reanalyzed by incorporating the other degree of freedom and the
entropy perturbation. Being interested in its implication to
inflation, here we will gather the general relevant results. In an
independent work \cite{Matsuda:2009np}, starting with more general
kinetic terms and more scalar fields, the evolution of the
``perturbed expansion rate'' was calculated for generalized gravity
theory using the techniques invented by \cite{Wands:2000dp}. We will
mainly follow the conventions and notations in \cite{Ji:2009yw}. For
instance, the signature of metric is $(-+++)$, and we will take
\begin{equation}\label{convetion}
M_p^{-2}=8\pi G,~~~~R=6(2H^2+\dot{H}),~~~~F=\frac{\partial}{\partial R}f(\varphi,R),~~~~E=\frac{2HF+\dot{F}}{F^\frac{3}{2}}.
\end{equation}
All of the general results have appeared in \cite{Ji:2009yw},
partly mixing with some special models. Nevertheless, it is still
helpful to put them orderly in this section.

First of all, let us define the slow-roll parameters:
\begin{eqnarray}\label{slroll-para}
\nonumber &&\epsilon_1=\frac{\dot{H}}{H^2},~~~~\epsilon_2=\frac{\ddot{H}}{H\dot{H}},~~~~\eta_1=\frac{\ddot{\varphi}}{H\dot{\varphi}},\\
\nonumber &&\eta_2=\frac{\dddot{\varphi}}{H\ddot{\varphi}},~~~~\delta_1=\frac{\dot{F}}{HF},~~~~\delta_2=\frac{\dot{E}}{HE},\\
&&\delta_3=\frac{\ddot{F}}{H\dot{F}},~~~~\delta_4=\frac{\ddot{E}}{H\dot{E}},~~~~\delta_6=\frac{\dddot{E}}{H\ddot{E}}~.
\end{eqnarray}
Be careful with the notation and sign difference between the
slow-roll parameters here and those in most literature. The
slow-roll conditions are met if the absolute values of these
parameters are much smaller than unity. Under the slow-roll
conditions, the background equations can be approximately written as
\begin{eqnarray}\label{Friedmann}
\nonumber &&V-\frac{1}{2}f+3H^2F\simeq0,\\
\nonumber &&\dot{\varphi}^2+2\dot{H}F-H\dot{F}\simeq0,\\
&&3H\dot{\varphi}-\frac{1}{2}f_{,\varphi}+V_{,\varphi}\simeq0.
\end{eqnarray}
These equations and the slow-roll conditions also result in the
following useful relations
\begin{eqnarray}
\nonumber &&\dot{E}=-\frac{\dot{\varphi}^2}{F^\frac{3}{2}},~~~~\ddot{E}=\frac{3\dot{F}\dot{\varphi}^2-4F\dot{\varphi}\ddot{\varphi}}{2F^\frac{5}{2}},~~~~\delta_2\simeq\epsilon_1-\frac{1}{2}\delta_1,\\
&&\delta_4\simeq2\eta_1-\frac{3}{2}\delta_1,~~~~\delta_6\simeq\eta_1-\frac{5}{2}\delta_1+\frac{3\delta_1\delta_3-\delta_1\eta_1-4\eta_1\eta_2}{3\delta_1-4\eta_1}.
\end{eqnarray}

In the longitudinal gauge, the
Friedmann-Lema\^{i}tre-Robertson-Walker (FLRW) metric with scalar
type perturbations is given by
\begin{equation}
ds^2=-(1+2\phi)dt^2+a^2(1-2\psi)\delta_{ij}dx^{i}dx^{j}.
\end{equation}
For generalized gravity, usually $\phi\neq\psi$, so we will have
two degrees of freedom after eliminating the inflaton fluctuation
$\delta\varphi$. The perturbed Einstein equations will give us two
coupled second-order differential equations for ($\phi$, $\psi$).
So we say there are two dynamical degrees of freedom. But this
pair of variables can be traded for ($\mathcal{R}$, $\mathcal{S}$)
or ($v_{\mathcal{R}}$, $v_{\mathcal{S}}$) or ($u_{\mathcal{R}}$,
$u_{\mathcal{S}}$) by the following relations:
\begin{eqnarray}\label{RSvu}
\nonumber \mathcal{R}&=&\frac{1}{2}(\phi+\psi)+\frac{2HF+\dot{F}}{2F\dot{\varphi}^2+3\dot{F}^2}\left[F(\dot{\phi}+\dot{\psi})+\frac{1}{2}(2HF+\dot{F})(\phi+\psi)\right],\\
\nonumber \mathcal{S}&=&\frac{\dot{F}}{\dot{\varphi}}\sqrt{\frac{3}{2F}}\left[\frac{F(2HF+\dot{F})}{2F\dot{\varphi}^2+3\dot{F}^2}(\dot{\phi}+\dot{\psi})+\frac{(2HF+\dot{F})^2}{4F\dot{\varphi}^2+6\dot{F}^2}(\phi+\psi)+\frac{2HF+\dot{F}}{2\dot{F}}(\phi-\psi)\right],\\
\nonumber v_{\mathcal{R}}&=&\frac{a\sqrt{2F(2F\dot{\varphi}^2+3\dot{F}^2)}}{2HF+\dot{F}}\mathcal{R},~~~~v_{\mathcal{S}}=\frac{a\sqrt{2F(2F\dot{\varphi}^2+3\dot{F}^2)}}{2HF+\dot{F}}\mathcal{S},\\
u_{\mathcal{R}}&=&\frac{F^\frac{3}{2}}{\sqrt{4F\dot{\varphi}^2+6\dot{F}^2}}(\phi+\psi),~~~~u_{\mathcal{S}}=\frac{a\sqrt{F(2F\dot{\varphi}^2+3\dot{F}^2)}}{\sqrt{3}(2HF+\dot{F})}\mathcal{S}.
\end{eqnarray}
We have chosen the normalization for $\mathcal{S}$ so that
$\mathcal{P}_{\mathcal{R}\ast}=\mathcal{P}_{\mathcal{S}\ast}$ when
perturbations cross the Hubble horizon, as will be given by equation
(\ref{spectra-def1}). In reference \cite{Ji:2009yw}, $\mathcal{R}$
is interpreted as the curvature perturbation while $\mathcal{S}$ is
interpreted as the entropy perturbation. $v_{\mathcal{R}}$ and
$v_{\mathcal{S}}$ are the corresponding canonical variables. The
interpretation of $u_{\mathcal{R}}$ and $u_{\mathcal{S}}$ is less
clear, but are defined for our convenience, and one may think
$u_{\mathcal{R}}$ as as something akin to the canonical momenta (not
exactly). In terms of them, the evolution equations of perturbations
take the form
\begin{eqnarray}\label{evol-u}
\nonumber u''_{\mathcal{R} k}+k^2u_{\mathcal{R} k}+m_{\mathcal{R}}^2a^2u_{\mathcal{R} k}+\beta u_{\mathcal{S}k}&=&0,\\
u''_{\mathcal{S}k}+k^2u_{\mathcal{S}k}+m_{\mathcal{S}}^2a^2u_{\mathcal{S}k}+\alpha k^2u_{\mathcal{R} k}&=&0,
\end{eqnarray}
whose coefficients
\begin{eqnarray}\label{coe}
\nonumber &&\frac{m_{\mathcal{R}}^2}{H^2}\simeq2\epsilon_1-\eta_1,~~~~\beta\simeq\sign(\dot{\varphi})aH\sqrt{\delta_1-2\epsilon_1},\\
\nonumber &&\alpha\simeq\sign(\dot{\varphi})\frac{2}{3}aH\sqrt{\delta_1-2\epsilon_1},\\
&&\frac{m_{\mathcal{S}}^2}{H^2}\simeq\frac{5}{2}\delta_1-5\epsilon_1+\frac{F}{3H^2F_{,R}}+\frac{\dot{F}F_{,\varphi}}{2H^2F_{,R}\dot{\varphi}}-6.
\end{eqnarray}

As is well known, couple is trouble. This also applies to the
coupled equations (\ref{evol-u}). To make our analysis simple, for
perturbations inside the Hubble horizon, we will always disregard
the coupling terms controlled by $\beta$ and $\alpha$ (decoupled
approximation). This enables us to get a rough estimation but also
induces some uncertainties. Imposing an appropriate quantized
initial condition at $k\gg aH$, we find an analytical solution
under the decoupled approximation,
\begin{eqnarray}\label{sol-u}
\nonumber u_{\mathcal{R}k}=-\frac{1}{4k^{\frac{3}{2}}}e^{i(\nu_1-\frac{1}{2})\frac{\pi}{2}}\sqrt{-\pi k\tau}H^{(1)}_{\nu_1}(-k\tau)\hat{e}_{\mathcal{R} k},&&\text{with}~\nu_1^2=\frac{1}{4}-\frac{m_{\mathcal{R}}^2}{H^2},\\
u_{\mathcal{S}k}=-\frac{1}{2\sqrt{6k}}e^{i(\nu_2-\frac{3}{2})\frac{\pi}{2}}\sqrt{-\pi k\tau}H^{(1)}_{\nu_2}(-k\tau)\hat{e}_{\mathcal{S}k},&&\text{with}~\nu_2^2=\frac{1}{4}-\frac{m_{\mathcal{S}}^2}{H^2}.
\end{eqnarray}
Here $\{\hat{e}_{\mathcal{R}k},\hat{e}_{\mathcal{S}k}\}$ is the
orthonormal basis
\begin{equation}
\langle\hat{e}_{\alpha k},\hat{e}_{\beta k'}\rangle=\delta_{\alpha\beta}\delta(k-k'),~~~~\alpha,\beta=\mathcal{R},\mathcal{S}.
\end{equation}
The normalization of Fourier modes $\mathcal{R}_k$ and
$\mathcal{S}_k$ is exhibited by equation (\ref{Fourier-def1}) in
appendix \ref{app-3pt}. If $m_{\mathcal{R}}^2/H^2\simeq0$ and
$m_{\mathcal{S}}^2/H^2\simeq-2$, then the power spectra at the
horizon-crossing $k=aH$ are nearly scale-invariant,
\begin{eqnarray}\label{spectra-def1}
\nonumber \mathcal{P}_{\mathcal{R}\ast}&=&\frac{k^3}{2\pi^2}|\mathcal{R}_{k\ast}|^2\simeq\left.\frac{H^4}{4\pi^2\dot{\varphi}^2}\right|_{\ast},\\
\mathcal{P}_{\mathcal{S}\ast}&=&\frac{k^3}{2\pi^2}|\mathcal{S}_{k\ast}|^2\simeq\left.\frac{H^4}{4\pi^2\dot{\varphi}^2}\right|_{\ast}.
\end{eqnarray}
The condition $m_{\mathcal{R}}^2/H^2\simeq0$ is trivial due to the
slow-roll conditions. But $m_{\mathcal{S}}^2/H^2\simeq-2$ puts a
nontrivial constraint on viable models. Throughout this paper, an
asterisk means the quantities take their horizon-crossing value.
Since we have neglected the coupling between curvature perturbation
and entropy perturbation inside the horizon, their cross-correlation
is negligible at the Hubble-crossing,
\begin{equation}\label{spectra-def2}
\mathcal{P}_{\mathcal{C}\ast}=\frac{k^3}{2\pi^2}\langle\mathcal{R}_{k\ast},\mathcal{S}_{k\ast}\rangle\simeq0.
\end{equation}
When crossing the horizon, the spectral indices are
\begin{equation}\label{nRnS0}
n_{\mathcal{R}\ast}-1=n_{\mathcal{S}\ast}-1=4\epsilon_{1\ast}-2\eta_{1\ast}.
\end{equation}

Unlike the single-field inflation in Einstein gravity, the entropy
perturbation and curvature perturbation are not conserved even well
outside the horizon $k\ll aH$. It is more convenient to follow their
evolution in terms of ($\mathcal{R}$, $\mathcal{S}$). If
$m_{\mathcal{S}}^2/H^2\simeq-2$, we have
$\ddot{\mathcal{S}}_k/(H\dot{\mathcal{S}}_k)\sim\mathcal{O}(\epsilon)$
and hence
\begin{eqnarray}\label{trans-eqn}
\nonumber \dot{\mathcal{S}}_k=\mu_{\mathcal{S}}H\mathcal{S}_k,&&\mu_{\mathcal{S}}=-\frac{1}{3}\left(\frac{m_{\mathcal{S}}^2}{H^2}+2+\epsilon_1-\frac{9}{4}\delta_1-3\delta_2+\frac{3}{2}\delta_4\right),\\
\dot{\mathcal{R}}_k=\mu_{\mathcal{R}}H\mathcal{S}_k,&&\mu_{\mathcal{R}}=\frac{\sign(\dot{\varphi})}{\delta_1}\sqrt{\frac{2(\delta_1-2\epsilon_1)}{3}}\left(2\eta_1-\frac{5}{2}\delta_1-\delta_4\right).
\end{eqnarray}
Taking $\mu_{\mathcal{S}}$ and $\mu_{\mathcal{R}}$ as constants
approximately, its analytical solution reads
\begin{eqnarray}\label{sol-trans}
\nonumber &&\mathcal{S}_k=\mathcal{S}_{k\ast}\exp\left(\int_{t_{\ast}}^t\mu_{\mathcal{S}}Hdt\right)=\mathcal{S}_{k\ast}e^{\mu_{\mathcal{S}}(N_{\ast}-N)},\\
&&\mathcal{R}_k-\mathcal{R}_{k\ast}=\int_{t_{\ast}}^t\mu_{\mathcal{R}}H\mathcal{S}_kdt=\frac{\mu_{\mathcal{R}}}{\mu_{\mathcal{S}}}\mathcal{S}_{k\ast}\left[e^{\mu_{\mathcal{S}}(N_{\ast}-N)}-1\right],
\end{eqnarray}
in which $N=\ln[a_{end}/a(t)]$ stands for the e-folding number from
time $t$ to the end of inflation.

As a result, on the super-hubble scale, the power spectra are
\begin{eqnarray}\label{spectra transition}
\nonumber &&\mathcal{P}_{\mathcal{R}}\simeq\mathcal{P}_{\mathcal{R}\ast}+\mathcal{P}_{\mathcal{S}\ast}\frac{\mu_{\mathcal{R}}^2}{\mu_{\mathcal{S}}^2}\left[e^{\mu_{\mathcal{S}}(N_{\ast}-N)}-1\right]^2,\\
\nonumber &&\mathcal{P}_{\mathcal{S}}\simeq\mathcal{P}_{\mathcal{S}\ast}e^{2\mu_{\mathcal{S}}(N_{\ast}-N)},\\
&&\mathcal{P}_{\mathcal{C}}\simeq\mathcal{P}_{\mathcal{S}\ast}\frac{\mu_{\mathcal{R}}}{\mu_{\mathcal{S}}}e^{\mu_{\mathcal{S}}(N_{\ast}-N)}\left[e^{\mu_{\mathcal{S}}(N_{\ast}-N)}-1\right].
\end{eqnarray}
Their spectral indices are
\begin{eqnarray}
\nonumber n_{\mathcal{R}}-1&=&n_{\mathcal{S}\ast}-1-\frac{2\mu_{\mathcal{R}}^2\mu_{\mathcal{S}}e^{\mu_{\mathcal{S}}(N_{\ast}-N)}\left[e^{\mu_{\mathcal{S}}(N_{\ast}-N)}-1\right]}{\mu_{\mathcal{S}}^2+\mu_{\mathcal{R}}^2\left[e^{\mu_{\mathcal{S}}(N_{\ast}-N)}-1\right]^2},\\
\nonumber n_{\mathcal{S}}-1&=&n_{\mathcal{S}\ast}-1-2\mu_{\mathcal{S}},\\
n_{\mathcal{C}}-1&=&n_{\mathcal{S}\ast}-1-\frac{\mu_{\mathcal{S}}\left[2e^{\mu_{\mathcal{S}}(N_{\ast}-N)}-1\right]}{e^{\mu_{\mathcal{S}}(N_{\ast}-N)}-1}.
\end{eqnarray}
We have defined the entropy-to-curvature ratio in \cite{Ji:2009yw}
\begin{equation}\label{ratio-def1}
r_{\mathcal{S}}=\frac{\mathcal{P}_{\mathcal{S}}}{\mathcal{P}_{\mathcal{R}}}.
\end{equation}

The tensor type perturbation is conserved outside the horizon. Its
power spectrum is relatively simple \cite{Hwang:2005hb}
\begin{equation}\label{spectra-def3}
\mathcal{P}_{T}\simeq\frac{2H^2}{\pi^2F}=\mathcal{P}_{\mathcal{S}\ast}(8\delta_1-16\epsilon_1),
\end{equation}
with a spectral index
\begin{equation}\label{nT}
n_{T}\simeq\frac{2\dot{H}}{H^2}-\frac{\dot{F}}{HF}=2\epsilon_1-\delta_1.
\end{equation}
Here we have used a normalization different from
\cite{Hwang:2005hb,Ji:2009yw} to accommodate to the WMAP5 convention
\cite{Komatsu:2008hk}.

The above results hold generally for slow-roll inflation in
generalized $f(\varphi,R)$ gravity with $F>0$, as long as the
entropy perturbation is non-vanishing and nearly scale-invariant.
For details of derivation and explanations, one can refer to
\cite{Ji:2009yw}.

It proves helpful to utilize also the entropy-curvature
correlation angle $\Delta$ \cite{Wands:2002bn} and the
tensor-to-scalar ratio $r_{T}$
\begin{equation}\label{ratio-def2}
\cos\Delta=\frac{\mathcal{P}_{\mathcal{C}}}{\sqrt{\mathcal{P}_{\mathcal{R}}\mathcal{P}_{\mathcal{S}}}},~~~~r_{T}=\frac{\mathcal{P}_{T}}{\mathcal{P}_{\mathcal{R}}},
\end{equation}
as well as a general transfer matrix
\begin{equation}\label{trans-matrix}
\left(
\begin{array}{c}
\mathcal{R}\\
\mathcal{S}
\end{array}
\right)=\left(
\begin{array}{cc}
1&T_{\mathcal{R}\mathcal{S}}\\
0&T_{\mathcal{S}\mathcal{S}}
\end{array}
\right)\left(\begin{array}{c}
\mathcal{R}\\
\mathcal{S}
\end{array}
\right)_{\ast}.
\end{equation}
In fact, $\cos\Delta$ is nothing else but the correlation
coefficient introduced in \cite{Langlois:1999dw}.

\section{Non-Gaussianity}\label{sect-Gauss}
The primordial non-Gaussianity has attracted a lot of attention
during recent years.\footnote{For a partial list, see
\cite{Komatsu:2001rj,Chen:2006nt,Maldacena:2002vr,Li:2008gg,Komatsu:2002db,Li:2008qc,Chen:2008wn,Huang:2008ze,Li:2008qv,Gao:2008dt,Li:2008jn,Xue:2008mk,Li:2008fma,Huang:2008qf,Huang:2008rj,Li:2008tw,Huang:2008bg,Huang:2008zj,Ling:2008xd,Moroi:2008nn,Tzirakis:2008qy,Gao:2009gd,Lyth:2005du,Lyth:2005fi,Zaballa:2006pv,Cogollo:2008bi,Rodriguez:2008hy}
and references therein. We also recommend \cite{Komatsu:2009kd} as
an up-to-date brief overview.} To judge whether we can get some
interesting large non-Gaussianity before expanding actions to the
third order and calculating the three-point correlation functions,
we can make some semi-quantitative estimates using the results of
single-field inflation in Einstein gravity theory.

We start with the calculation of three-point correlation of
curvature perturbations $\mathcal{R}$, by virtue of
\eqref{trans-matrix},
\begin{eqnarray}\label{<RRR>}
\nonumber &&\langle\mathcal{R}(\mathbf{k}_1)\mathcal{R}(\mathbf{k}_2)\mathcal{R}(\mathbf{k}_3)\rangle\\
\nonumber &=&\langle(\mathcal{R}(\mathbf{k}_1)_\ast+T_{\mathcal{RS}}\mathcal{S}(\mathbf{k}_1)_\ast)(\mathcal{R}(\mathbf{k}_2)_\ast+T_{\mathcal{RS}}\mathcal{S}(\mathbf{k}_2)_\ast)(\mathcal{R}(\mathbf{k}_3)_\ast+T_{\mathcal{RS}}\mathcal{S}(\mathbf{k}_3)_\ast)\rangle\\
\nonumber &=&\langle\mathcal{S}(\mathbf{k}_1)_\ast\mathcal{S}(\mathbf{k}_2)_\ast\mathcal{S}(\mathbf{k}_3)_\ast\rangle T_\mathcal{RS}^3\\
\nonumber &&+\left[\langle\mathcal{R}(\mathbf{k}_3)_\ast\mathcal{S}(\mathbf{k}_1)_\ast\mathcal{S}(\mathbf{k}_2)_\ast\rangle+\langle\mathcal{R}(\mathbf{k}_2)_\ast\mathcal{S}(\mathbf{k}_3)_\ast\mathcal{S}(\mathbf{k}_1)_\ast\rangle+\langle\mathcal{R}(\mathbf{k}_1)_\ast\mathcal{S}(\mathbf{k}_2)_\ast\mathcal{S}(\mathbf{k}_3)_\ast\rangle\right]T_\mathcal{RS}^2\\
\nonumber &&+\left[\langle\mathcal{R}(\mathbf{k}_2)_\ast\mathcal{R}(\mathbf{k}_3)_\ast\mathcal{S}(\mathbf{k}_1)_\ast\rangle+\langle\mathcal{R}(\mathbf{k}_1)_\ast\mathcal{R}(\mathbf{k}_3)_\ast\mathcal{S}(\mathbf{k}_2)_\ast\rangle+\langle\mathcal{R}(\mathbf{k}_1)_\ast\mathcal{R}(\mathbf{k}_2)_\ast\mathcal{S}(\mathbf{k}_3)_\ast\rangle\right]T_\mathcal{RS}\\
&&+\langle\mathcal{R}(\mathbf{k}_1)_\ast\mathcal{R}(\mathbf{k}_2)_\ast\mathcal{R}(\mathbf{k}_3)_\ast\rangle,
\end{eqnarray}
where an asterisk means the quantities are calculated at the time of
horizon-crossing $k=aH$. In the above equation, we assumed the
linear evolution of $\mathcal{R}_k$ and $\mathcal{S}_k$ outside the
horizon. Although this assumption is good enough for our
semi-quantitative analysis, in a more accurate treatment, one should
consider the nonlinear effects. There are two sources of nonlinear
effects outside the horizon: the $\ddot{S}_k$ term neglected in
equation \eqref{trans-eqn}; the time dependence of
$\mu_{\mathcal{S}}$ and $\mu_{\mathcal{R}}$.

As we have mentioned, the curvature perturbation and the entropy
perturbation are coupled. But, under our approximation, their
coupling inside the horizon will not be taken into consideration in
the estimation of magnitude. Because all of these quantities are
calculated at horizon-crossing, we can treat the adiabatic and
entropy perturbations independently. Using the single-field
relations \cite{Komatsu:2001rj,Chen:2006nt}, \footnote{The notation
of the momentum modes of the perturbations in \cite{Chen:2006nt} are
different from here by some different choice of normalization in
Fourier expansions. See appendix \ref{app-3pt} for details.}
\begin{eqnarray}\label{<RRR>-fNL<RR>}
\nonumber \langle\mathcal{R}(\mathbf{k}_1)_\ast\mathcal{R}(\mathbf{k}_2)_\ast\mathcal{R}(\mathbf{k}_3)_\ast\rangle&=&(2\pi)^{5/2}\delta^{(3)}\left(\sum_i\mathbf{k}_i\right)\left[-\frac{3}{10}f_{NL\ast}^{\mathcal{R}}\left(\mathcal{P}_{k\ast}^\mathcal{R}\right)^2\right]\frac{\sum_ik_i^3}{\prod_ik_i^3},\\
\langle\mathcal{S}(\mathbf{k}_1)_\ast\mathcal{S}(\mathbf{k}_2)_\ast\mathcal{S}(\mathbf{k}_3)_\ast\rangle&=&(2\pi)^{5/2}\delta^{(3)}\left(\sum_i\mathbf{k}_i\right)\left[-\frac{3}{10}f_{NL\ast}^{\mathcal{S}}\left(\mathcal{P}_{k\ast}^\mathcal{S}\right)^2\right]\frac{\sum_ik_i^3}{\prod_ik_i^3}.
\end{eqnarray}
The power spectra on the right hand side are given by
\eqref{spectra-def1} and related by \eqref{ratio-def1}, while the
nonlinear parameters $f_{NL\ast}^\mathcal{R}$ and
$f_{NL\ast}^\mathcal{S}$ can be estimated independently by the
single-field results. Hence the summation involving these two pure
three-point correlations can be easily written into a compact
form,
\begin{eqnarray}
\nonumber &&\langle\mathcal{S}(\mathbf{k}_1)_\ast\mathcal{S}(\mathbf{k}_2)_\ast\mathcal{S}(\mathbf{k}_3)_\ast\rangle T_\mathcal{RS}^3+\langle\mathcal{R}(\mathbf{k}_1)_\ast\mathcal{R}(\mathbf{k}_2)_\ast\mathcal{R}(\mathbf{k}_3)_\ast\rangle\\
&=&(2\pi)^{5/2}\delta^{(3)}\left(\sum_i\mathbf{k}_i\right)\left(-\frac{3}{10}\right)\left(f_{NL\ast}^{\mathcal{R}}+r_{\mathcal{S}\ast}^2T_{\mathcal{RS}}^3f_{NL\ast}^{\mathcal{S}}\right)\left(\mathcal{P}_{k\ast}^\mathcal{R}\right)^2\frac{\sum_ik_i^3}{\prod_ik_i^3}.
\end{eqnarray}

The contributions of terms like $\langle\mathcal{RSS}\rangle$ and
$\langle\mathcal{RRS}\rangle$ will be a little trouble before we
know the exact forms of third order action and perform careful
calculations. Here we cannot determine the value of these terms
for generic configuration, but in appendix \ref{app-3pt} there is
an estimation for the local shape. We find there the three-point
correlations involving both $\mathcal{R}$ and $\mathcal{S}$ are
proportional to the two-point cross-correlation
$\mathcal{P}_{\mathcal{C}}$, whose initial value is negligible at
the time of horizon-crossing under our
approximation.\footnote{However, as argued in \cite{Ji:2009yw}
along the line of \cite{Byrnes:2006fr}, the cross-correlation may
not be negligible,
$\mathcal{P}_{\mathcal{C}\ast}/\mathcal{P}_{\mathcal{S}\ast}\sim\mathcal{O}(\alpha)\sim\mathcal{O}(\beta)$,
were the coupling terms taken into consideration. Of course, even
if we take them into account, due to the $\alpha$ or $\beta$
suppression, the dominant contribution is still given by
$\langle\mathcal{R}\mathcal{R}\mathcal{R}\rangle$ and
$\langle\mathcal{S}\mathcal{S}\mathcal{S}\rangle$ terms. So our
approximation captures the leading order contributions.} These
proportional relations rely on the locality of the shape, although
it is possible that they could be generalized to other shapes by
incorporating the $\mathbf{k}$ dependence of
$f_{NL}^{\mathcal{R}}$ and $f_{NL}^{\mathcal{S}}$. Strictly
speaking, we have to reevaluate their contributions seriously when
going beyond the local limit. But, lacking of a solid proof, we
will still set
$\langle\mathcal{R_{\ast}S_{\ast}S_{\ast}}\rangle=\langle\mathcal{R_{\ast}R_{\ast}S_{\ast}}\rangle=0$
for all shapes, which will give us some satisfactory results in
semi-quantitative estimation. In the previous section, we made a
decoupled approximation for linear perturbations inside the
horizon. The assumption here is just a nonlinear generalization of
that linear one. After this assumption, we get on the
super-horizon scale,
\begin{equation}\label{generic-3pt}
\langle\mathcal{R}(\mathbf{k}_1)\mathcal{R}(\mathbf{k}_2)\mathcal{R}(\mathbf{k}_3)\rangle=(2\pi)^{5/2}\delta^{(3)}\left(\sum_i\mathbf{k}_i\right)\left(-\frac{3}{10}\right)(\mathcal{P}_{k\ast}^\mathcal{R})^2\frac{\sum_ik_i^3}{\prod_ik_i^3}\left(f_{NL\ast}^{\mathcal{R}}+r_{\mathcal{S}\ast}^2T_{\mathcal{RS}}^3f_{NL\ast}^{\mathcal{S}}\right).
\end{equation}

At the same time, the left hand side of \eqref{generic-3pt} can be
converted into
\begin{equation}\label{consist-rel}
\langle\mathcal{R}(\mathbf{k}_1)\mathcal{R}(\mathbf{k}_2)\mathcal{R}(\mathbf{k}_3)\rangle=(2\pi)^{5/2}\delta^{(3)}\left(\sum_i\mathbf{k}_i\right)\left[-\frac{3}{10}f_{NL}^{\mathcal{R}}\left(\mathcal{P}_{k}^\mathcal{R}\right)^2\right]\frac{\sum_ik_i^3}{\prod_ik_i^3},
\end{equation}
where the spectrum can be related with the one at horizon-crossing
by \eqref{spectra transition}, as
\begin{equation}\label{spectrum-trans}
\mathcal{P}_\mathcal{R}=\mathcal{P}_{\mathcal{R}\ast}+T_\mathcal{RS}^2\mathcal{P}_{\mathcal{S}\ast}=\left(1+r_{\mathcal{S}\ast}T_\mathcal{RS}^2\right)\mathcal{P}_{\mathcal{R}\ast}.
\end{equation}
Comparing \eqref{generic-3pt}, \eqref{consist-rel} and
\eqref{spectrum-trans}, we finally get the nonlinear parameter
$f_{NL}$ of curvature perturbation on super-Hubble scale,
especially at the end of inflation, expressed by some parameters at
horizon-crossing,
\begin{equation}\label{generic-fNL}
f_{NL}=\frac{f_{NL\ast}^{\mathcal{R}}+r_{\mathcal{S}\ast}^2T_{\mathcal{RS}}^3f_{NL\ast}^{\mathcal{S}}}{(1+r_{\mathcal{S}\ast}T_\mathcal{RS}^2)^2}
\end{equation}
Here $f_{NL\ast}^{\mathcal{R}}$ and $f_{NL\ast}^{\mathcal{S}}$ are
computed at $k=aH$. In our approximation, the curvature and entropy
perturbations are evolving independently before that time, so the
nonlinear parameters for them at the Hubble-exit can be estimated
with the independent single-field results
\cite{Maldacena:2002vr,Li:2008gg},
\begin{eqnarray}\label{single-fNL}
\nonumber f_{NL\ast}^{\mathcal{R}}&=&-\frac{5}{12}[n_{\mathcal{R}\ast}-1+f(\mathbf{k}_1,\mathbf{k}_2,\mathbf{k}_3)n_{T\ast}],\\
f_{NL\ast}^{\mathcal{S}}&=&-\frac{5}{12\xi}\left[n_{\mathcal{S}\ast}-1-\frac{2\dot{\xi}}{H\xi}+f(\mathbf{k}_1,\mathbf{k}_2,\mathbf{k}_3)n_{T\ast}\right],
\end{eqnarray}
where $f(\mathbf{k}_1,\mathbf{k}_2,\mathbf{k}_3)$ is a factor of
momentum configuration, with maximum $5/6$ in equilateral limit
and minimum $0$ in local limit \cite{Maldacena:2002vr}. $\xi$ is a
factor related to the normalization of $\mathcal{S}$.
Corresponding to our normalization (\ref{RSvu}), it is
\begin{equation}
\xi=\frac{\dot{F}}{\dot{\varphi}}\sqrt{\frac{3}{2F}},~~~~\frac{\dot{\xi}}{H\xi}=\delta_3-\eta_1-\frac{1}{2}\delta_1.
\end{equation}
In our model, so far we do not know the relation between the entropy
perturbation during inflation and the one at the matter-radiation
decoupling. So there is an ambiguity in the normalization of entropy
perturbation. In literature of two-field inflation, a convenient
normalization is usually chosen so that
$\mathcal{P}_{\mathcal{R}\ast}=\mathcal{P}_{\mathcal{S}\ast}$ at the
Hubble-exit. We follow the same normalization. But one should
realize that the it is $\mathcal{S}/\xi$ rather than $\mathcal{S}$
that satisfies the simplest form of the consistency relation.
Therefore the second consistency relation (\ref{single-fNL}) takes a
relatively more complicated form. Since
$n_{\mathcal{R}\ast}-1=n_{\mathcal{S}\ast}-1$ and
$r_{\mathcal{S}\ast}=1$ to the leading order, then we have a
simplified estimation of \eqref{generic-fNL} as
\begin{eqnarray}
\nonumber f_{NL}&=&\frac{f_{NL\ast}^{\mathcal{R}}(\xi+T_{\mathcal{RS}}^3)}{\xi(1+T_\mathcal{RS}^2)^2}+\frac{5\dot{\xi}T_{\mathcal{RS}}^3}{6H\xi^2(1+T_\mathcal{RS}^2)^2}\\
&=&-\frac{5(\xi+T_{\mathcal{RS}}^3)}{12\xi(1+T_\mathcal{RS}^2)^2}[n_{\mathcal{R}\ast}-1+f(\mathbf{k}_1,\mathbf{k}_2,\mathbf{k}_3)n_{T\ast}]+\frac{5\dot{\xi}T_{\mathcal{RS}}^3}{6H\xi^2(1+T_\mathcal{RS}^2)^2}.
\end{eqnarray}

For nowadays observation, the most relevant results are its values
in the local and the equilateral limits:
\begin{eqnarray}\label{fNL-final}
\nonumber  f_{NL}^{local}&=&-\frac{5(\xi+T_\mathcal{RS}^3)}{12\xi(1+T_\mathcal{RS}^2)^2}(n_{\mathcal{R}\ast}-1)+\frac{5\dot{\xi}T_{\mathcal{RS}}^3}{6H\xi^2(1+T_\mathcal{RS}^2)^2},\\
f_{NL}^{equil}&=&-\frac{5(\xi+T_\mathcal{RS}^3)}{12\xi(1+T_\mathcal{RS}^2)^2}\left(n_{\mathcal{R}\ast}-1+\frac{5}{6}n_{T\ast}\right)+\frac{5\dot{\xi}T_{\mathcal{RS}}^3}{6H\xi^2(1+T_\mathcal{RS}^2)^2}.
\end{eqnarray}
All the parameters involved in this formula can be expressed by
the slow-roll parameters and e-folding number, as in the previous
section. We will evaluate the results for specific models given
below in section \ref{sect-quadratic} and \ref{sect-quartic}.

\section{$1/R$ Correction to Inflation}\label{sect-Rinverse}
With the above results at hand, it is straightforward to study the
inflation model (\ref{action-gRin}), where a inflaton-dependent
$1/R$ correction is included. The steps are parallel to those in
\cite{Ji:2009yw}. Remember in \cite{Ji:2009yw} a special model
with a inflaton-dependent $R^2$ term was considered. But it turned
out the inflaton is rolling up its potential in that model. It is
a rather tricky problem to terminate inflation in ``rolling-up''
models. So it would be interesting to get a ``rolling-down'' model
in $f(\varphi,R)$ gravity. The model we are going to study has
this quality, to which we will return at the end of this section.

Comparing (\ref{action-gRin}) with (\ref{action-gg}), it reads
directly,
\begin{equation}
f(\varphi,R)=M_p^2R+\frac{g(\varphi)}{R},~~~~F=M_p^2-\frac{g}{R^2},
\end{equation}
then we obtain the simplified background equations
\begin{eqnarray}
\label{simp-Friedmann1}&&V-3M_p^2H^2-\frac{g}{16H^2}\simeq0,\\
\label{simp-Friedmann2}&&\dot{\varphi}^2+2\dot{H}F-H\dot{F}\simeq0,\\
\label{simp-eom-varphi}&&3H\dot{\varphi}\simeq\frac{g_{,\varphi}}{24H^2}-V_{,\varphi}.
\end{eqnarray}
The solutions for equation (\ref{simp-Friedmann1}) are simply
\begin{equation}\label{rho}
H^2=\frac{1}{3M_p^2}\rho(\varphi)=\frac{V\pm\sqrt{V^2-\frac{3}{4}gM_p^2}}{6M_p^2}.
\end{equation}
We will always take the positive solution (the one with upper sign
``$+$'') by virtue of the fact $H^2>0$.

We get the following relations:
\begin{eqnarray}
\nonumber &&F\simeq M_p^2-\frac{gM_p^4}{16\rho^2},\\
\nonumber &&\rho_{,\varphi}\dot{\varphi}\simeq6M_p^2H\dot{H}=2H\rho\epsilon_1,\\
\nonumber &&\rho_{,\varphi}\ddot{\varphi}+\rho_{,\varphi\varphi}\dot{\varphi}^2\simeq6M_p^2(H\ddot{H}+\dot{H}^2),\\
\nonumber &&3H\ddot{\varphi}+3\dot{H}\dot{\varphi}\simeq\left(\frac{g_{,\varphi}M_p^2}{8\rho}-V_{,\varphi}\right)_{,\varphi}\dot{\varphi},\\
&&3H\dddot{\varphi}+6\dot{H}\ddot{\varphi}+3\ddot{H}\dot{\varphi}\simeq\left(\frac{g_{,\varphi}M_p^2}{8\rho}-V_{,\varphi}\right)_{,\varphi}\ddot{\varphi}+\left(\frac{g_{,\varphi}M_p^2}{8\rho}-V_{,\varphi}\right)_{,\varphi\varphi}\dot{\varphi}^2.
\end{eqnarray}
The slow-roll parameters (\ref{slroll-para}) can be expressed in
terms of $g$ and $V$ and their derivatives with respect to
$\varphi$,
\begin{eqnarray}
\nonumber &&\epsilon_1=\frac{\dot{H}}{H^2}\simeq\frac{\rho_{,\varphi}M_p^2}{2\rho^2}\left(\frac{g_{,\varphi}M_p^2}{8\rho}-V_{,\varphi}\right),\\
\nonumber &&\eta_1=\frac{\ddot{\varphi}}{H\dot{\varphi}}\simeq-\epsilon_1+\frac{M_p^2}{\rho}\left(\frac{g_{,\varphi}M_p^2}{8\rho}-V_{,\varphi}\right)_{,\varphi},\\
\nonumber &&\delta_1=\frac{\dot{F}}{HF}\simeq-\frac{2\epsilon_1\rho^3M_p^2}{\rho_{,\varphi}(16\rho^2-gM_p^2)}\left(\frac{g}{\rho^2}\right)_{,\varphi},~~~~\delta_2=\frac{\dot{E}}{HE}\simeq\epsilon_1-\frac{1}{2}\delta_1,\\
\nonumber &&\delta_3=\frac{\ddot{F}}{H\dot{F}}=\eta_1+\frac{2\epsilon\rho(g/\rho^2)_{,\varphi\varphi}}{\rho_{,\varphi}(g/\rho^2)_{,\varphi}},~~~~\delta_4=\frac{\ddot{E}}{H\dot{E}}\simeq2\eta_1-\frac{3}{2}\delta_1,\\
\nonumber &&\epsilon_2=\frac{\ddot{H}}{H\dot{H}}\simeq\eta_1-\epsilon_1+\frac{2\epsilon\rho\rho_{,\varphi\varphi}}{\rho_{,\varphi}^2},\\
\nonumber &&\eta_2=\frac{\dddot{\varphi}}{H\ddot{\varphi}}\simeq\eta_1-\epsilon_1-\frac{\epsilon_1\epsilon_2}{\eta_1}+\frac{2\epsilon_1M_p^2}{\eta_1\rho_{,\varphi}}\left(\frac{g_{,\varphi}M_p^2}{8\rho}-V_{,\varphi}\right)_{,\varphi\varphi},\\
&&\delta_6=\frac{\dddot{E}}{H\ddot{E}}\simeq\eta_1-\frac{5}{2}\delta_1+\frac{3\delta_1\delta_3-\delta_1\eta_1-4\eta_1\eta_2}{3\delta_1-4\eta_1}.
\end{eqnarray}
The ``mass squared'' for entropy perturbation reduces to
\begin{equation}
\frac{m_{\mathcal{S}}^2}{H^2}\simeq\frac{5}{2}\delta_1-6\epsilon_1-8+\frac{32\rho^2}{gM_p^2}\left(1+\frac{3}{2}\epsilon_1\right)+\frac{3g_{,\varphi}M_p^4}{16g}\left(\frac{g}{\rho^2}\right)_{,\varphi}\left(1+\frac{1}{2}\epsilon_1\right).
\end{equation}
To calculate non-Gaussianities, we also need
\begin{equation}
\xi=-\left(\frac{gM_p^3}{8\rho^2}\right)_{,\varphi}\sqrt{\frac{6\rho^2}{16\rho^2-gM_p^2}}.
\end{equation}

In order to move on, we should specify the potential $V(\varphi)$
and the coupling $g(\varphi)$. This is the task of the coming two
sections. Here we should mention the condition to make the power
spectra scale-invariant. There nontrivial condition
$m_{\mathcal{S}}^2/H^2\simeq-2$ is translated now to the requirement
$32\rho^2/(gM_p^2)\sim6$. According to (\ref{rho}), this requirement
is easy to satisfy if we choose
\begin{equation}\label{gV}
g=\frac{4V^2-M_p^4V_{,\varphi\varphi}^2}{3M_p^2}.
\end{equation}
We will take this choice in the subsequent sections. The
$V_{,\varphi\varphi}$ term in the numerator is necessary,
otherwise one would find $\dot{F}=0$ and $\mu_{\mathcal{R}}$ is
divergent. Generally we have $4V^2\gg M_p^4V_{,\varphi\varphi}^2$,
then from (\ref{simp-eom-varphi}) it is not hard to get
\begin{equation}
3H\dot{\varphi}\simeq-\frac{1}{3}V_{,\varphi}.
\end{equation}
In our following specific examples $V=\frac{1}{2}m^2\varphi^2$ or
$V=\lambda\varphi^4$ ($\lambda>0$), so $V_{,\varphi}\dot{\varphi}<0$
and the inflaton is rolling down its potential as promised. We also
have $F>0$, so the formalism developed in \cite{Ji:2009yw} is
applicable here.

\section{Quadratic Potential: $V(\varphi)=\frac{1}{2}m^2\varphi^2$}\label{sect-quadratic}
In this case, the action takes the form
\begin{equation}\label{action-Rin2}
S=\int d^4x\sqrt{-g}\left[\frac{1}{2}M_p^2R+\frac{m^4(\varphi^4-M_p^4)}{6M_p^2R}-\frac{1}{2}g^{\alpha\beta}\partial_{\alpha}\varphi\partial_{\beta}\varphi-\frac{1}{2}m^2\varphi^2\right].
\end{equation}
When the scalar field $\varphi$ fades out, this action recovers the
gravitational part of action (\ref{action-Rin}) if $\mu^4=m^4/3$.
But as we will see at the end of this section, this is not the case
because $\mu^4\ll m^4/3$.

\begin{figure}
\begin{center}
\includegraphics[width=0.5\textwidth]{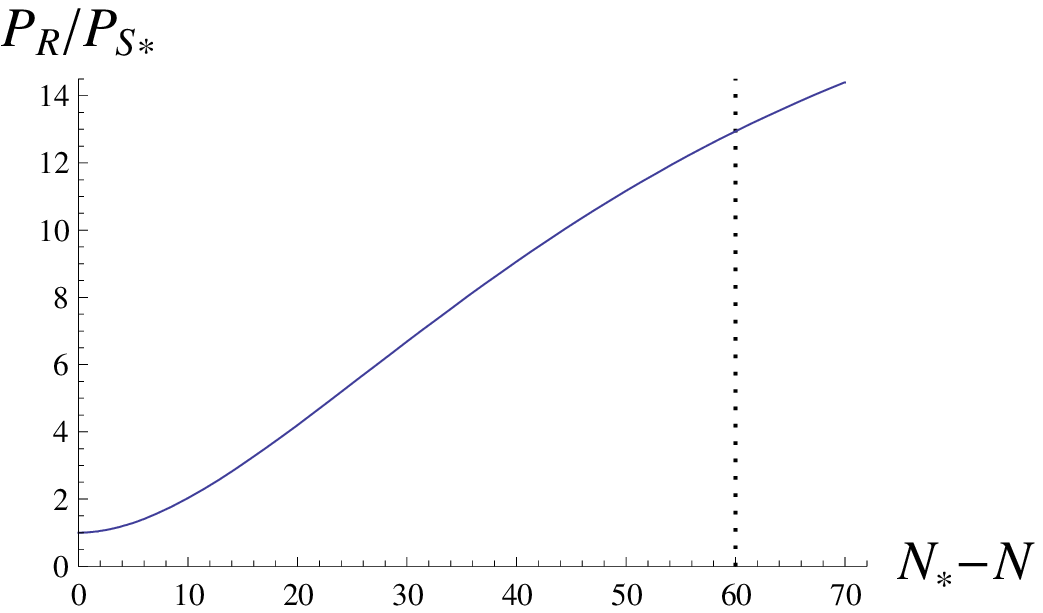}\\
\includegraphics[width=0.5\textwidth]{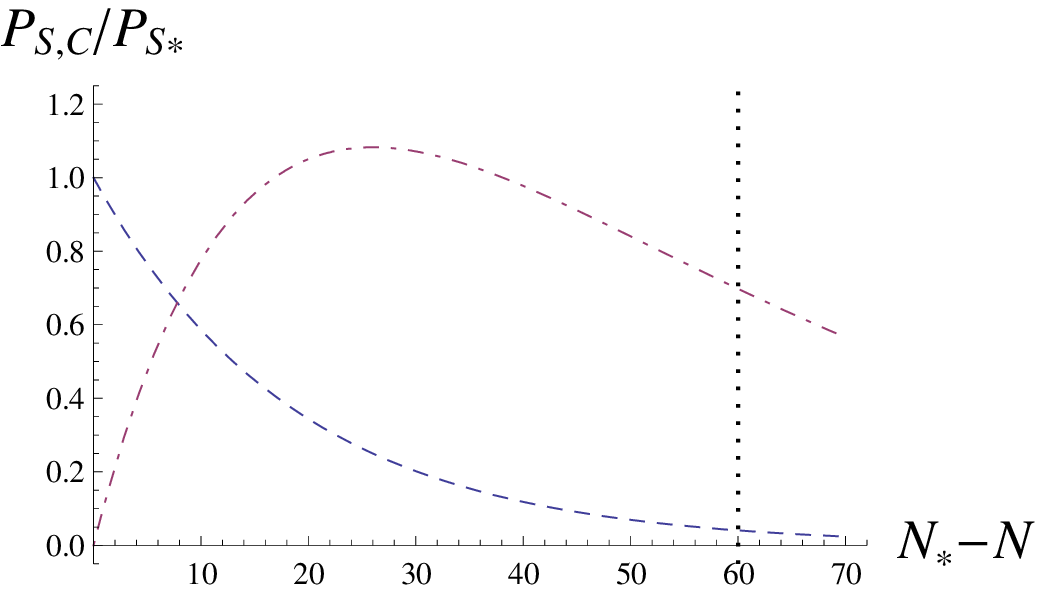}\\
\includegraphics[width=0.5\textwidth]{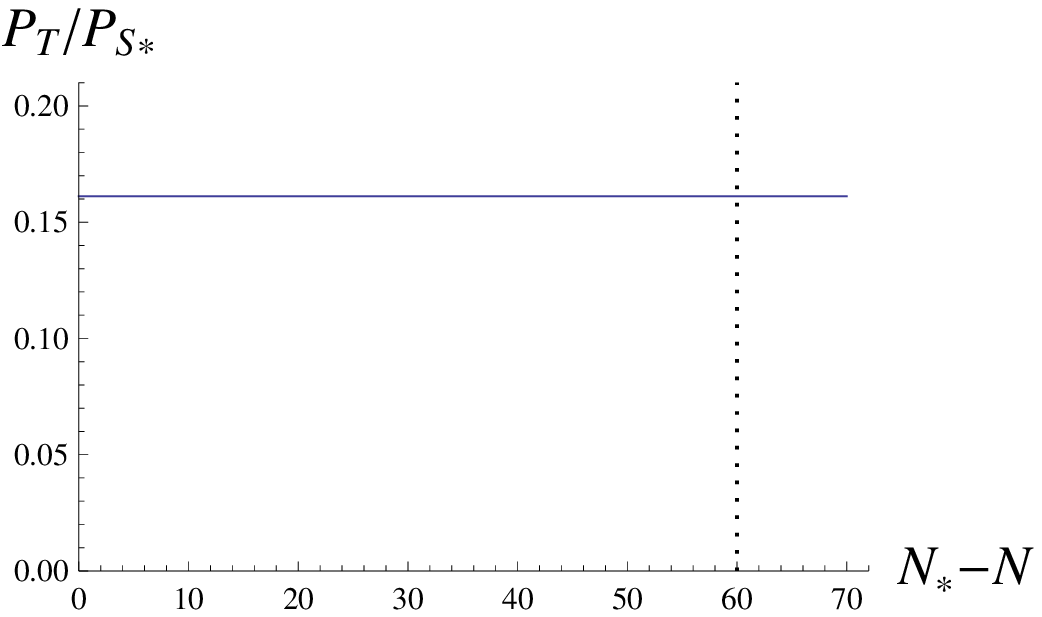}\\
\end{center}
\caption{\textbf{The evolutions of power spectra with respect to
e-folding number $N_{\ast}-N$ after crossing the horizon. This
figure is drawn according to the model with action
(\ref{action-Rin2}). From top to bottom: curvature power spectrum
$\mathcal{P}_{\mathcal{R}}$, entropy power spectrum
$\mathcal{P}_{\mathcal{S}}$ (dashed blue line) and cross-correlation
power spectrum $\mathcal{P}_{\mathcal{C}}$ (dot-dashed purple line),
tensor power spectrum $\mathcal{P}_{T}$. All of the power spectra
are normalized by $\mathcal{P}_{\mathcal{S\ast}}$, the entropy power
spectrum at horizon-crossing. The vertical dotted black lines
correspond to $N_{\ast}-N=60$.}}\label{fig-spectra2}
\end{figure}

\begin{figure}
\includegraphics[width=0.5\textwidth]{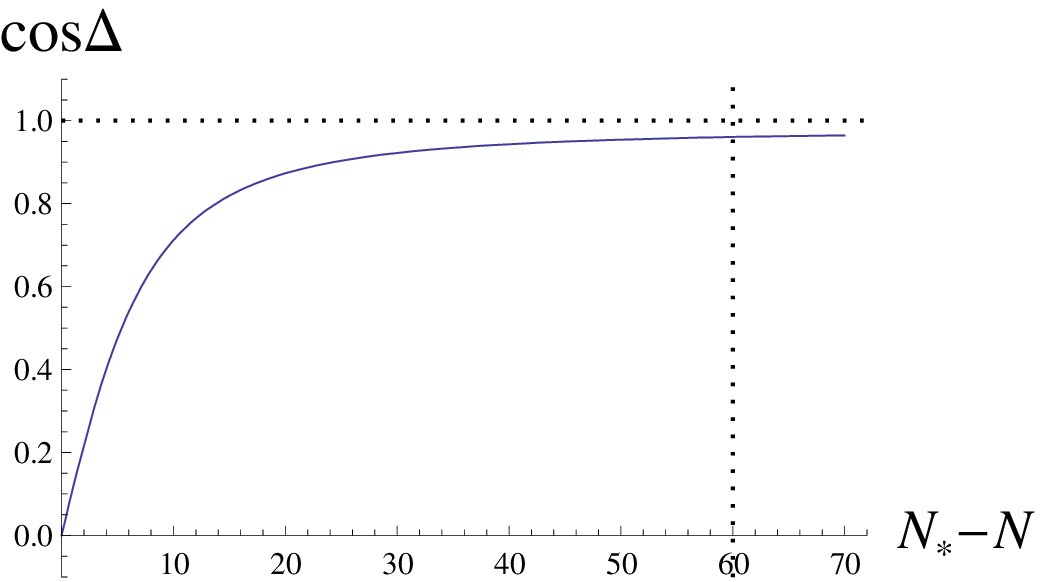}\\
\includegraphics[width=0.5\textwidth]{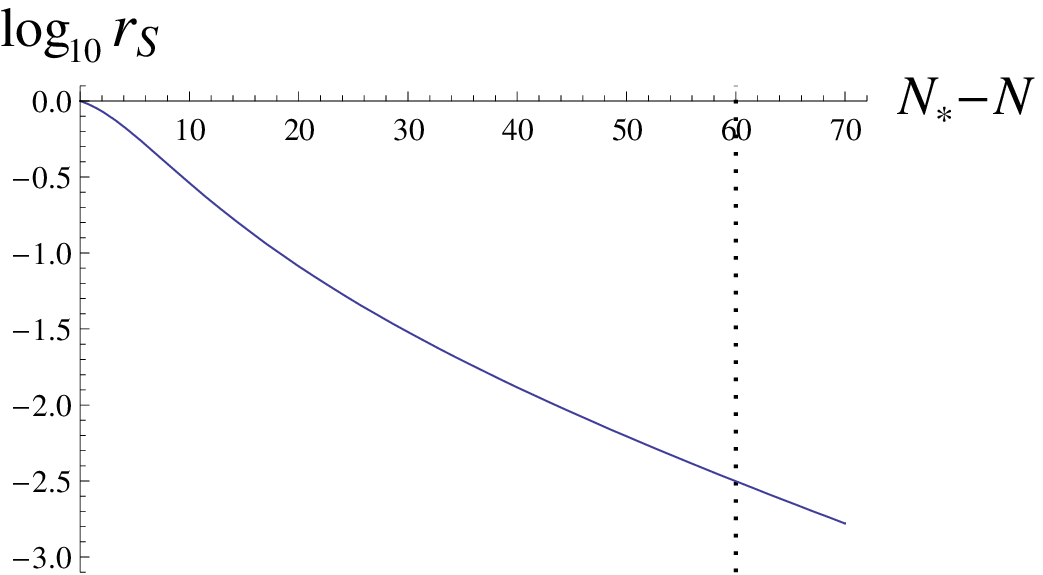}\\
\includegraphics[width=0.5\textwidth]{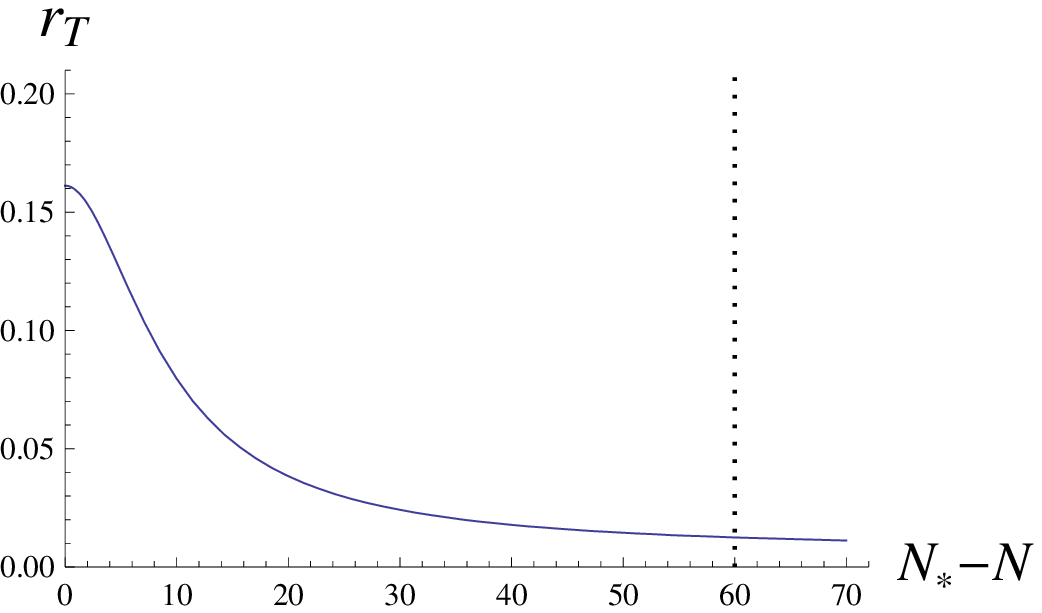}\\
\caption{\textbf{The evolutions of correlation coefficient
$\cos\Delta$ (upper graph), entropy-to-curvature ratio
$r_{\mathcal{S}}$ (its logarithm, middle graph) and tensor-to-scalar
ratio $r_{T}$ (lower graph) with respect to e-folding number
$N_{\ast}-N$ after crossing the horizon. This figure is drawn
according to the model with action (\ref{action-Rin2}). The vertical
dotted black lines correspond to $N_{\ast}-N=60$. The horizontal
dotted black line corresponds to $\cos\Delta=1$, that is, the
totally correlated situation.}}\label{fig-ratio2}
\end{figure}

\begin{figure}
\includegraphics[width=0.5\textwidth]{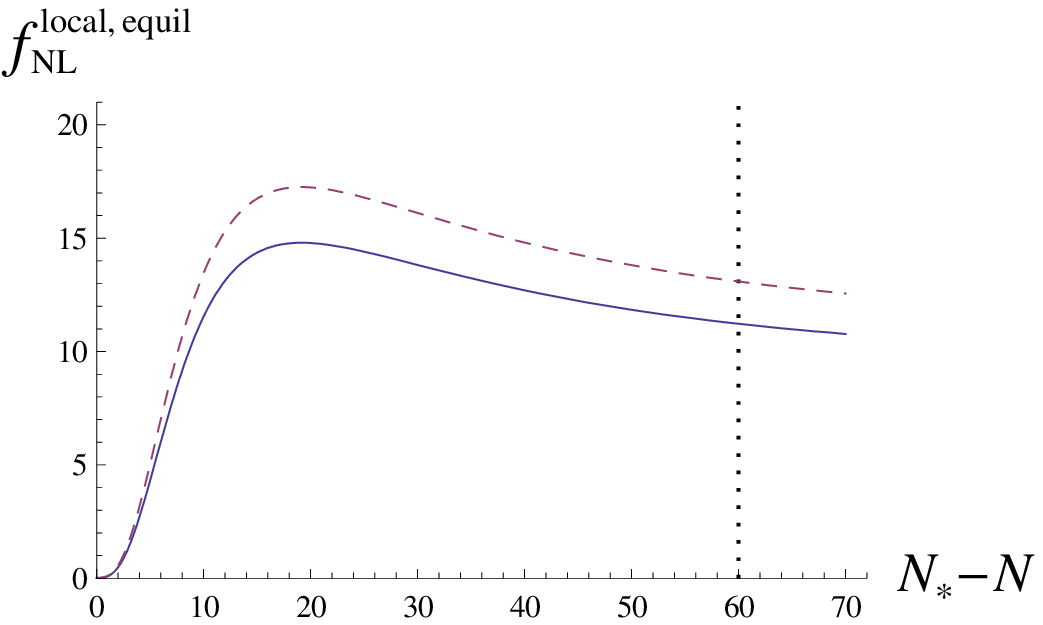}\\
\caption{\textbf{The evolutions of nonlinear parameters of curvature
perturbation with respect to e-folding number $N_{\ast}-N$ after
crossing the horizon. The solid blue line corresponds to the local
limit value $f_{NL}^{local}$, while the dashed purple line depicts
the nonlinear parameter of equilateral shape $f_{NL}^{equil}$. The
vertical dotted black line corresponds to $N_{\ast}-N=60$. This
figure is drawn according to the model with action
(\ref{action-Rin2}).}}\label{fig-fNL2}
\end{figure}

For later convenience, let us define a new notation
\begin{equation}
\gamma=\frac{M_p^2}{\varphi^2}.
\end{equation}
This notation is also useful in the next section. In the present
case, one can prove
\begin{equation}
\epsilon_1+\eta_1\simeq-\frac{4M_p^2(3M_p^4+\varphi^4)}{3(M_p^2+\varphi^2)^3}.
\end{equation}
According to this expression, the condition $\gamma\ll1$ is
necessary in order to satisfy the slow-roll conditions. As will be
clear below, this is also the sufficient condition to meet the
slow-roll conditions. So we can conclude that this model describes
the large field inflation.

All of the slow-roll parameters can be expressed in terms of
$\gamma$ to the leading order as
\begin{eqnarray}\label{slroll-Rin2}
\nonumber &&\epsilon_1=\frac{\dot{H}}{H^2}\simeq-\frac{4}{3}\gamma,~~~~\eta_1=\frac{\ddot{\varphi}}{H\dot{\varphi}}\simeq4\gamma^2,~~~~\delta_1=\frac{\dot{F}}{HF}\simeq\frac{8}{3}\gamma^2,\\
\nonumber &&\delta_2=\frac{\dot{E}}{HE}\simeq-\frac{4}{3}\gamma,~~~~\delta_3=\frac{\ddot{F}}{H\dot{F}}\simeq4\gamma,~~~~\delta_4=\frac{\ddot{E}}{H\dot{E}}\simeq4\gamma^2,\\
&&\epsilon_2=\frac{\ddot{H}}{H\dot{H}}\simeq\frac{8}{3}\gamma^2,~~~~\eta_2=\frac{\dddot{\varphi}}{H\ddot{\varphi}}\simeq4\gamma,~~~~\delta_6=\frac{\dddot{E}}{H\ddot{E}}\simeq4\gamma.
\end{eqnarray}
The coefficients (\ref{coe}) are
\begin{eqnarray}
\nonumber \frac{m_{\mathcal{R}}^2}{H^2}\simeq-\frac{8}{3}\gamma,&&\beta\simeq-2\sqrt{\frac{2\gamma}{3}},\\
\alpha\simeq-\frac{4}{3}\sqrt{\frac{2\gamma}{3}},&&\frac{m_{\mathcal{S}}^2}{H^2}\simeq-2+8\gamma.
\end{eqnarray}

When the perturbations cross the horizon,
\begin{eqnarray}
\nonumber &&\mathcal{P}_{\mathcal{R}\ast}=\mathcal{P}_{\mathcal{S}\ast}=\frac{3m^2\varphi_{\ast}^4}{256\pi^2M_p^6},\\
\nonumber &&r_{\mathcal{S}\ast}=1,~~~~n_{\mathcal{R}\ast}-1=n_{\mathcal{S}\ast}-1=-\frac{16}{3}\gamma_{\ast},\\
\nonumber &&\mu_{\mathcal{S}\ast}=-\frac{32}{3}\gamma_{\ast},~~~~\mu_{\mathcal{R}\ast}=\frac{4}{3}\sqrt{\gamma_{\ast}},\\
&&\xi_{\ast}=\sqrt{2}\gamma_{\ast}^{\frac{3}{2}},~~~~\left.\frac{\dot{\xi}}{H\xi}\right|_{\ast}=4\gamma_{\ast}.
\end{eqnarray}

If we use the horizon-crossing value to estimate $\mu_{\mathcal{S}}$
and $\mu_{\mathcal{R}}$ outside the horizon, then at the end of
inflation ($N=0$),
\begin{eqnarray}
\nonumber &&\frac{\mathcal{P}_{\mathcal{R}}}{\mathcal{P}_{\mathcal{S}\ast}}=\frac{9}{64\gamma_{\ast}}\left(1-e^{-32N_{\ast}\gamma_{\ast}/9}\right)^2+1,\\
\nonumber &&\frac{\mathcal{P}_{\mathcal{S}}}{\mathcal{P}_{\mathcal{S}\ast}}=e^{-64N_{\ast}\gamma_{\ast}/9},~~~~\frac{\mathcal{P}_{T}}{\mathcal{P}_{\mathcal{S}\ast}}=\frac{64}{3}\gamma_{\ast},\\
\nonumber &&\frac{\mathcal{P}_{\mathcal{C}}}{\mathcal{P}_{\mathcal{S}\ast}}=\frac{3}{8\sqrt{\gamma_{\ast}}}e^{-32N_{\ast}\gamma_{\ast}/9}\left(1-e^{-32N_{\ast}\gamma_{\ast}/9}\right),\\
\nonumber &&n_{\mathcal{R}}-1=-\frac{16}{3}\gamma_{\ast},~~~~n_{\mathcal{S}}-1=\frac{16}{9}\gamma_{\ast},\\
&&n_{\mathcal{C}}-1=\frac{16}{3}\gamma_{\ast},~~~~n_{T}=-\frac{8}{3}\gamma_{\ast}.
\end{eqnarray}

To determine the parameter $\gamma_\ast$, we make use of the
observational constraint on curvature spectral index
$n_{\mathcal{R}}-1\simeq-0.04$, which gives approximately
$\gamma_\ast=3/400$. The results are shown in figures
\ref{fig-spectra2} and figure \ref{fig-ratio2}. In figure
\ref{fig-spectra2} we plot the evolution of spectra
$\mathcal{P}_{\mathcal{R}}$, $\mathcal{P}_{\mathcal{S}}$,
$\mathcal{P}_{\mathcal{C}}$ and $\mathcal{P}_{T}$, which are defined
in (\ref{spectra-def1}), (\ref{spectra-def2}) and
(\ref{spectra-def3}). When drawing the graph, we have normalized
them by $\mathcal{P}_{\mathcal{S}\ast}$. Figure \ref{fig-ratio2}
depicts the evolution of the correlation coefficient $\cos\Delta$,
the logarithm of entropy-to-curvature ratio $r_{\mathcal{S}}$ and
the tensor-to-scalar ratio $r_{T}$, defined by (\ref{ratio-def1})
and (\ref{ratio-def2}).

At the end of inflation, it can be seen from figure
\ref{fig-ratio2} that the entropy perturbation and the curvature
perturbation are almost totally correlated, and the
entropy-to-curvature ratio $r_{\mathcal{S}}$ is of order
$10^{-3}$. At first glance, the entropy-to-curvature ratio here
can be tested against WMAP5 constraint \cite{Komatsu:2008hk} as
done by \cite{Ji:2009yw}. However, this is a misleading game. What
WMAP5 constrained is the entropy perturbation
\begin{equation}
\mathcal{S}_{c,\gamma}=\frac{\delta\rho_c}{\rho_c}-\frac{3\delta\rho_\gamma}{4\rho_\gamma}
\end{equation}
between dark matter and radiation. In our model the entropy
perturbation \cite{Ji:2009yw}
\begin{equation}
\mathcal{S}\propto\frac{\delta\varphi}{\dot{\varphi}}-\frac{F}{\dot{F}}(\psi-\phi)
\end{equation}
which is related to the difference between the Newtonian potential
$\phi$ and the spatial curvature $\psi$. Firstly, the
normalization of $\mathcal{S}$ does not match to $\mathcal{S}$.
Second, it is unlikely that the two degrees of freedom in our
model will decay into radiation and dark matter respectively. Most
probably such an entropy perturbation would seed an anisotropic
stress or quadrupole moments of photons and neutrinos. Third, the
entropy mode may decay after inflation, which depends on the
detailed mechanism of reheating. Especially, the entropy
perturbation can be erased by thermal equilibrium of matter and
radiation before the creation of any non-zero conserved quantum
number \cite{Komatsu:2008hk,Weinberg:2004kr,Weinberg:2004kf}.

To estimate the non-Gaussianity, we calculate \eqref{fNL-final}
for the present case,
\begin{eqnarray}
\nonumber f_{NL}^{local}&=&\frac{40\left[2048\gamma_{\ast}^3+135\sqrt{2}\left(1-e^{-32N_{\ast}\gamma_{\ast}/9}\right)^3\right]}{9\left[64\gamma_{\ast}+9\left(1-e^{-32N_{\ast}\gamma_{\ast}/9}\right)^2\right]^2},\\
f_{NL}^{equil}&=&\frac{20\left[17408\gamma_{\ast}^3+945\sqrt{2}\left(1-e^{-32N_{\ast}\gamma_{\ast}/9}\right)^3\right]}{27\left[64\gamma_{\ast}+9\left(1-e^{-32N_{\ast}\gamma_{\ast}/9}\right)^2\right]^2}.
\end{eqnarray}
Once again we set $\gamma_\ast=3/400$, then the numerical result
gives the nonlinear parameter $f_{NL}^{local}$ and $f_{NL}^{equil}$
as functions of the e-folding number $N_\ast$. Both of them are
illustrated in figure \ref{fig-fNL2}. At the end of inflation, this
model will give the nonlinear parameters $f_{NL}^{local}\simeq11$
and $f_{NL}^{equil}\simeq13$.

Although all of the above predictions (or postdictions) are
consistent with observational data, this model suffers from a
serious problem, as we want to point out here. If we take
$n_{\mathcal{R}}-1=-0.04$, then the normalization of curvature power
spectrum $\mathcal{P}_{\mathcal{R}}\sim10^{-9}$ requires
$m^2/M_p^2\sim10^{-5}$. On the other hand, the smallness of
cosmological constant requires $\mu^2/M_p^2\sim10^{-121}$ in action
(\ref{action-gRin}). In other words, if one intends to use model
(\ref{action-Rin2}) to explain the comic microwave background (CMB)
anisotropy, the residual ``dark energy'' will be too large compared
with the observed value. For this reason, we conclude the model
(\ref{action-Rin2}) with a quadratic potential is unattractive. In
section \ref{sect-stab}, we will discuss another problem of it.

\section{Quartic Potential: $V(\varphi)=\lambda\varphi^4$}\label{sect-quartic}
Starting with the action
\begin{equation}\label{action-Rin4}
S=\int d^4x\sqrt{-g}\left[\frac{1}{2}M_p^2R+\frac{4\lambda^2\varphi^4(\varphi^4-36M_p^4)}{6M_p^2R}-\frac{1}{2}g^{\alpha\beta}\partial_{\alpha}\varphi\partial_{\beta}\varphi-\lambda\varphi^4\right],~~~~(\lambda>0)
\end{equation}
the treatment of this model is similar to the previous section, but
the result is more encouraging.

\begin{figure}
\begin{center}
\includegraphics[width=0.5\textwidth]{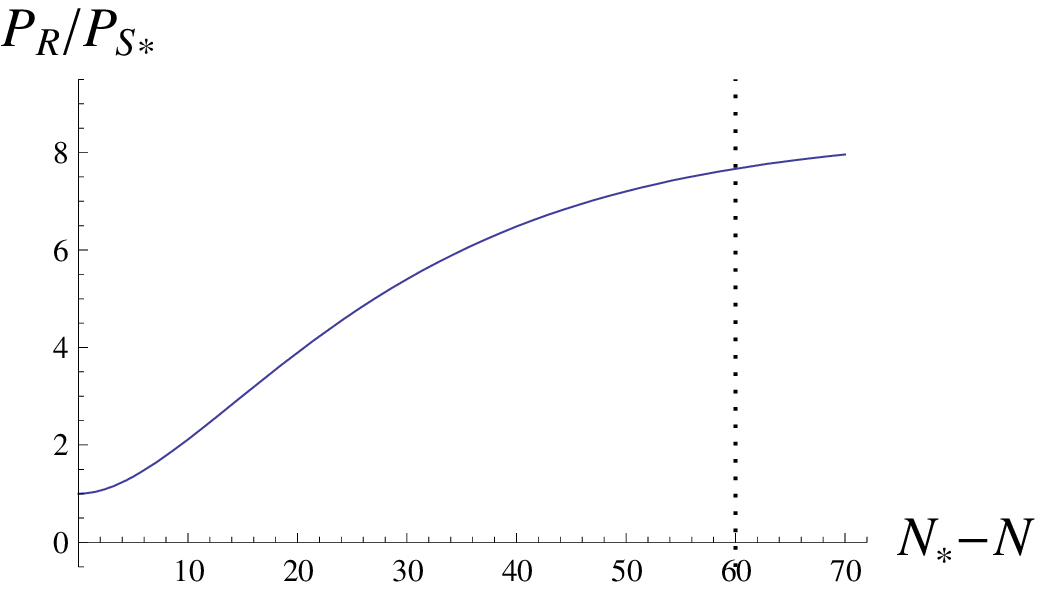}\\
\includegraphics[width=0.5\textwidth]{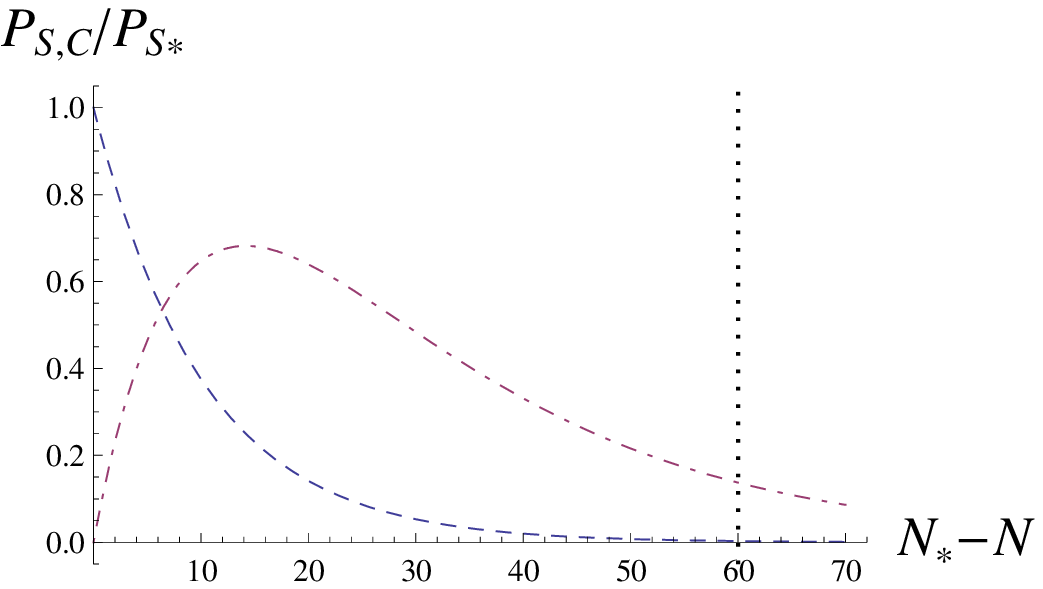}\\
\includegraphics[width=0.5\textwidth]{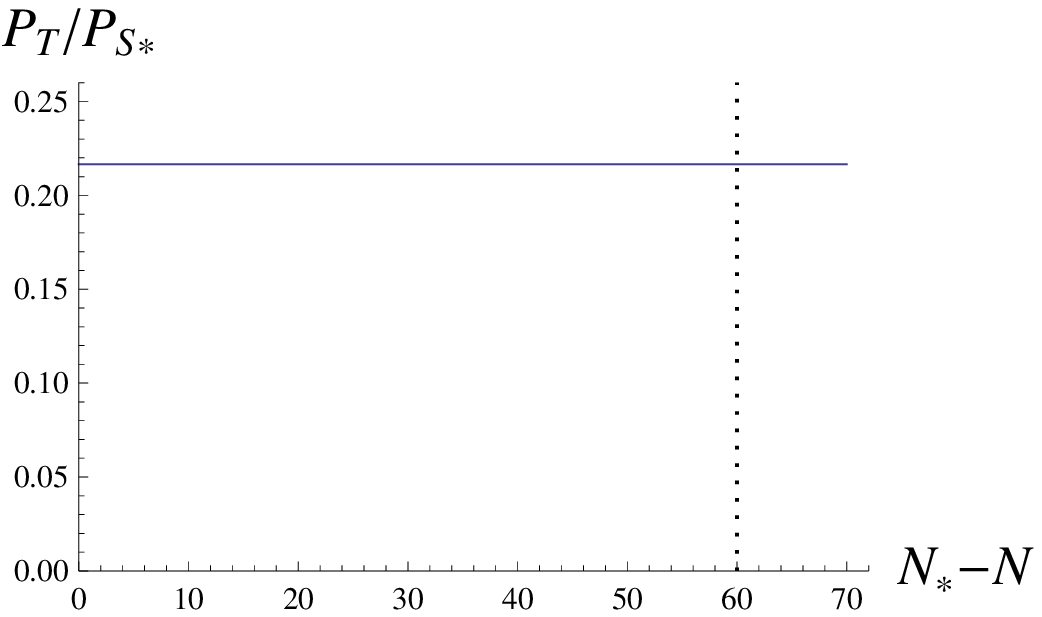}\\
\end{center}
\caption{\textbf{The evolutions of power spectra with respect to
e-folding number $N_{\ast}-N$ after crossing the horizon. We draw
this figure according to the model with action (\ref{action-Rin4}).
The upper graph depicts the evolution of curvature power spectrum
$\mathcal{P}_{\mathcal{R}}$. The middle graph depicts the evolution
curves of entropy power spectrum $\mathcal{P}_{\mathcal{S}}$ (dashed
blue line) and cross-correlation power spectrum
$\mathcal{P}_{\mathcal{C}}$ (dot-dashed purple line). The lower
corresponds to tensor power spectrum $\mathcal{P}_{T}$. All of the
power spectra are normalized by $\mathcal{P}_{\mathcal{S\ast}}$, the
entropy power spectrum at horizon-crossing. The vertical dotted
black lines correspond to $N_{\ast}-N=60$.}}\label{fig-spectra4}
\end{figure}

\begin{figure}
\includegraphics[width=0.5\textwidth]{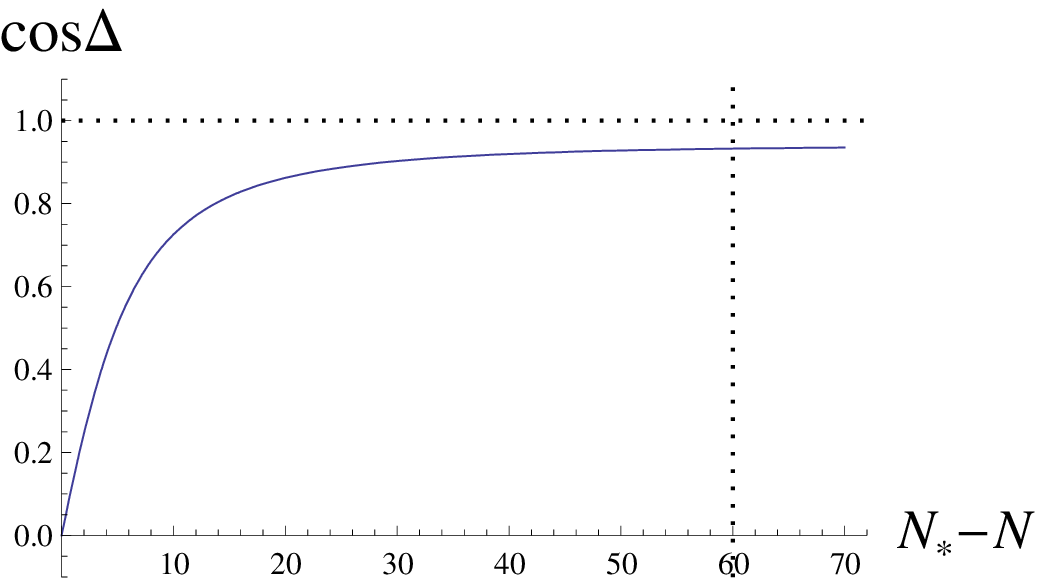}\\
\includegraphics[width=0.5\textwidth]{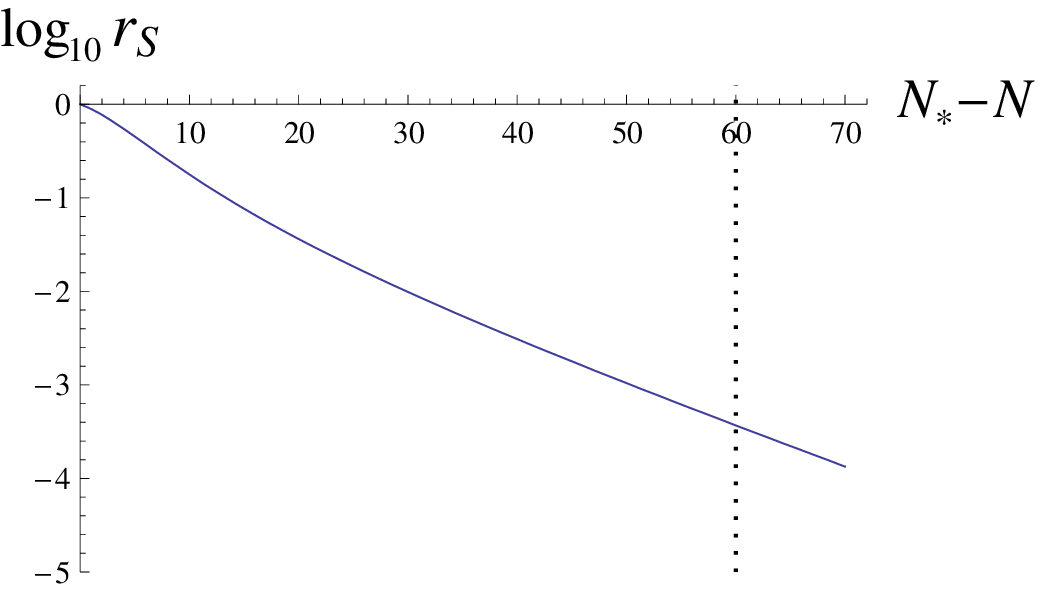}\\
\includegraphics[width=0.5\textwidth]{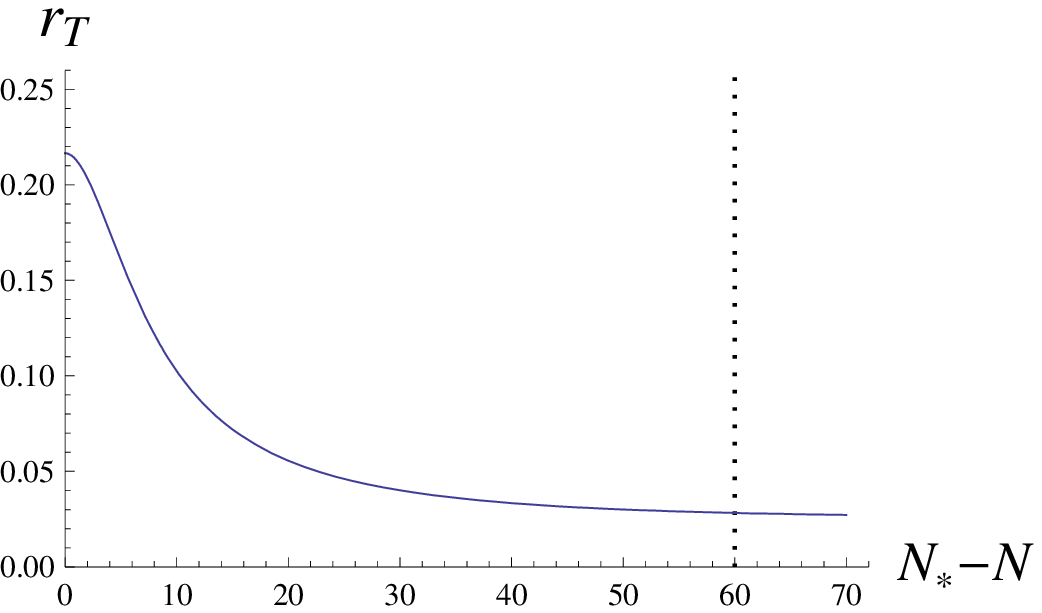}\\
\caption{\textbf{The evolutions of correlation coefficient
$\cos\Delta$ (upper graph), entropy-to-curvature ratio
$r_{\mathcal{S}}$ (its logarithm, middle graph) and tensor-to-scalar
ratio $r_{T}$ (lower graph) with respect to e-folding number
$N_{*}-N$ after crossing the horizon. This figure is drawn according
to the model with action (\ref{action-Rin4}). The vertical dotted
black lines correspond to $N_{*}-N=60$. The horizontal dotted black
line corresponds to $\cos\Delta=-1$, that is, the totally
anti-correlated situation.}}\label{fig-ratio4}
\end{figure}

\begin{figure}
\includegraphics[width=0.5\textwidth]{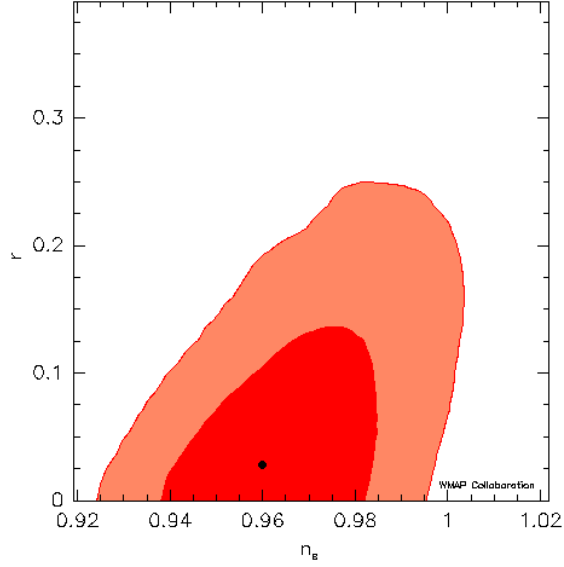}\\
\caption{\textbf{The black dot is the prediction of quartic model
(\ref{action-Rin4}) for $n_{\mathcal{R}}$ and $r_{T}$, where we have
set $\gamma_{\ast}=1/400$. It is consistent with the constraint from
WMAP5 + BAO (baryon acoustic oscillations) + SN (supernovae)
\cite{Komatsu:2008hk}.}}\label{ns-vs-r}
\end{figure}

\begin{figure}
\includegraphics[width=0.5\textwidth]{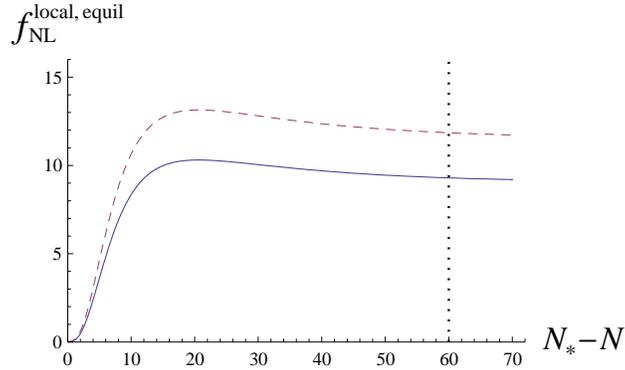}\\
\caption{\textbf{The evolutions of nonlinear parameters of curvature
perturbation with respect to e-folding number $N_{\ast}-N$ after
crossing the horizon. The solid blue curve corresponds to the local
limit value $f_{NL}^{local}$, while the dashed purple curve plots
the value in equilateral limit $f_{NL}^{equil}$. The vertical dotted
black line corresponds to $N_{\ast}-N=60$. This figure is drawn
according to the model with action
(\ref{action-Rin4}).}}\label{fig-fNL4}
\end{figure}

Again we find the necessary condition $\gamma=M_p^2/\varphi^2\ll1$
for slow-roll because of the relation
\begin{equation}
\epsilon_1+\eta_1\simeq-\frac{8M_p^2(72M_p^6+96M_p^4\varphi^2+16M_p^2\varphi^4+\varphi^6)}{3\varphi^2(6M_p^2+\varphi^2)^3}.
\end{equation}

To the leading order of $\gamma$, we write the slow-roll parameters
in the present case
\begin{eqnarray}\label{slroll-Rin4}
\nonumber &&\epsilon_1=\frac{\dot{H}}{H^2}\simeq-\frac{16}{3}\gamma,~~~~\eta_1=\frac{\ddot{\varphi}}{H\dot{\varphi}}\simeq-\frac{8}{3}\gamma,~~~~\delta_1=\frac{\dot{F}}{HF}\simeq32\gamma^2,\\
\nonumber &&\delta_2=\frac{\dot{E}}{HE}\simeq-\frac{16}{3}\gamma,~~~~\delta_3=\frac{\ddot{F}}{H\dot{F}}\simeq\frac{16}{3}\gamma,~~~~\delta_4=\frac{\ddot{E}}{H\dot{E}}\simeq-\frac{16}{3}\gamma,\\
&&\epsilon_2=\frac{\ddot{H}}{H\dot{H}}\simeq-\frac{16}{3}\gamma^2,~~~~\eta_2=\frac{\dddot{\varphi}}{H\ddot{\varphi}}\simeq-\frac{8}{3}\gamma,~~~~\delta_6=\frac{\dddot{E}}{H\ddot{E}}\simeq-\frac{16}{3}\gamma.
\end{eqnarray}
and the coefficients (\ref{coe}) for evolution equations,
\begin{eqnarray}
\nonumber \frac{m_{\mathcal{R}}^2}{H^2}\simeq-8\gamma,&&\beta\simeq-4\sqrt{\frac{2\gamma}{3}},\\
\alpha\simeq-\frac{8}{3}\sqrt{\frac{2\gamma}{3}},&&\frac{m_{\mathcal{S}}^2}{H^2}\simeq-2+56\gamma.
\end{eqnarray}

When the perturbations cross the horizon,
\begin{eqnarray}
\nonumber &&\mathcal{P}_{\mathcal{R}\ast}=\mathcal{P}_{\mathcal{S}\ast}=\frac{3\lambda\varphi_{\ast}^6}{512\pi^2M_p^6},\\
\nonumber &&r_{\mathcal{S}\ast}=1,~~~~n_{\mathcal{R}\ast}-1=n_{\mathcal{S}\ast}-1=-16\gamma_{\ast},\\
\nonumber &&\mu_{\mathcal{S}\ast}=-\frac{176}{9}\gamma_{\ast},~~~~\mu_{\mathcal{R}\ast}=\frac{8}{3}\sqrt{\gamma_{\ast}},\\
&&\xi_{\ast}=6\sqrt{2}\gamma_{\ast}^{\frac{3}{2}},~~~~\left.\frac{\dot{\xi}}{H\xi}\right|_{\ast}=8\gamma_{\ast}.
\end{eqnarray}

If we use the horizon-crossing value to estimate $\mu_{\mathcal{S}}$
and $\mu_{\mathcal{R}}$ outside the horizon, then at the end of
inflation ($N=0$), the power spectra and spectral indices are
\begin{eqnarray}
\nonumber &&\frac{\mathcal{P}_{\mathcal{R}}}{\mathcal{P}_{\mathcal{S}\ast}}=\frac{9}{484\gamma_{\ast}}\left(1-e^{-176N_{\ast}\gamma_{\ast}/9}\right)^2+1,\\
\nonumber &&\frac{\mathcal{P}_{\mathcal{S}}}{\mathcal{P}_{\mathcal{S}\ast}}=e^{-352N_{\ast}\gamma_{\ast}/9},~~~~\frac{\mathcal{P}_{T}}{\mathcal{P}_{\mathcal{S}\ast}}=\frac{256}{3}\gamma_{\ast},\\
\nonumber &&\frac{\mathcal{P}_{\mathcal{C}}}{\mathcal{P}_{\mathcal{S}\ast}}=\frac{3}{22\sqrt{\gamma_{\ast}}}e^{-176N_{\ast}\gamma_{\ast}/9}\left(1-e^{-176N_{\ast}\gamma_{\ast}/9}\right),\\
\nonumber &&n_{\mathcal{R}}-1=-16\gamma_{\ast},~~~~n_{\mathcal{S}}-1=\frac{208}{9}\gamma_{\ast},\\
&&n_{\mathcal{C}}-1=\frac{128}{3}\gamma_{\ast},~~~~n_{T}=-\frac{32}{3}\gamma_{\ast}.
\end{eqnarray}

From the horizon-crossing to the end of inflation, the power
spectrum of curvature perturbation has increased significantly. In
sharp contrast, the entropy perturbation drops down exponentially
with respect to $N_{\ast}-N$. The cross-correlation between them
takes a positive value, at first increasing in amplitude and then
decreasing. As we have promised, the tensor type perturbation is
invariant. These results are presented in figure
\ref{fig-spectra4}.

Now turn to figure \ref{fig-ratio4}. Look at the upper graph for
the evolution of correlation coefficient $\cos\Delta$. Under our
approximation, at the time of Hubble-crossing ($N_{\ast}-N=0$),
the curvature perturbation and the entropy perturbation are
uncorrelated. But the the subsequent evolution makes them almost
totally correlated at the end of inflation ($N_{\ast}-N=60$).

From the lower graph of figure \ref{fig-ratio4}, it is clear that
the tensor-to-scalar ratio is depressed greatly even at the
horizon-crossing time (compared with the $\lambda\varphi^4$
inflation in Einstein gravity). Let us take a closer look on this
point. Given the choice (\ref{gV}), at the time of
Hubble-crossing, the power spectrum for curvature perturbation and
that for tensor type perturbation can be written as
\begin{equation}\label{PRPT-Rin}
\mathcal{P}_{\mathcal{R}\ast}=\frac{3V^3}{32\pi^2M_p^6V_{,\varphi}^2},~~~~\mathcal{P}_{T\ast}=\frac{V}{2\pi^2M_p^2}.
\end{equation}
In contrast, the counterparts in Einstein gravity are given by
\begin{equation}\label{PRPT-Ein}
\mathcal{P}_{\mathcal{R}}|_{\mathrm{Einstein}}=\frac{V^3}{12\pi^2M_p^6V_{,\varphi}^2},~~~~\mathcal{P}_{T}|_{\mathrm{Einstein}}=\frac{2V}{3\pi^2M_p^2}.
\end{equation}
The difference between (\ref{PRPT-Rin}) and (\ref{PRPT-Ein})
explains the smallness of $r_{T}$ at the horizon-crossing in our
model. Outside the horizon, since the curvature perturbation is
increasing while the tensor type perturbation is conserved, the
value of $r_{T}$ becomes smaller and smaller. At the end of
inflation, we have $r_{T}\simeq0.03$. This is well inside the
constraint of WMAP5 \cite{Komatsu:2008hk}, as illustrated in
figure \ref{ns-vs-r}.

Words are needed here about the entropy-to-curvature ratio
$r_{\mathcal{S}}$, which is depicted in the middle graph of figure
\ref{fig-ratio4}. Thanks to the exponential decrease of entropy
perturbation, the value of $r_{\mathcal{S}}$ is of order $10^{-4}$
at the end of inflation, which is smaller than the WMAP5 upper
bound \cite{Komatsu:2008hk}. However, as we have emphasized in the
previous section, it is not quite reasonable to compare the
entropy perturbation here with that in WMAP5 result. A more
relevant constraint might come from the quadrupole moments of
neutrinos. Especially, since the entropy perturbation is very
small at the end of inflation, we can treat $\phi+\psi$ as almost
constant at that time. From equation (\ref{RSvu}), this gives
\begin{equation}
\psi-\phi=\frac{\dot{F}(2HF+\dot{F})}{2F\dot{\varphi}^2+3\dot{F}^2}(\phi+\psi)\propto\gamma_{\ast}\psi
\end{equation}
in our specific model. It is an interesting question to investigate
the implication of residual difference between spatial curvature and
Newtonian potential. But we do not pursue it furthermore in this
paper.

The non-Gaussian features can be studied as before. By virtue of
the new relations between the slow-roll parameters and
$\gamma_\ast$ we have
\begin{eqnarray}
\nonumber f_{NL}^{local}&=&\frac{220\left[21296\gamma_{\ast}^3+9\sqrt{2}\left(1-e^{-176N_{\ast}\gamma_{\ast}/9}\right)^3\right]}{3\left[484\gamma_{\ast}+9\left(1-e^{-176N_{\ast}\gamma_{\ast}/9}\right)^2\right]^2},\\
f_{NL}^{equil}&=&\frac{110\left[596288\gamma_{\ast}^3+207\sqrt{2}\left(1-e^{-176N_{\ast}\gamma_{\ast}/9}\right)^3\right]}{27\left[484\gamma_{\ast}+9\left(1-e^{-176N_{\ast}\gamma_{\ast}/9}\right)^2\right]^2}.
\end{eqnarray}
One can plot the evolution of $f_{NL}^{local}$ and $f_{NL}^{equil}$
with respect to $N_\ast-N$ in figure \ref{fig-fNL4}. At the end of
inflation, this model will give the nonlinear parameters
$f_{NL}^{local}\simeq9$ and $f_{NL}^{equil}\simeq12$. Of course,
these numbers are just results of semi-quantitative estimation. If
we take them seriously, we would like to compare them with
observational constraints \cite{Komatsu:2008hk}. They perfectly
satisfy the WMAP5 limit $-9<f_{NL}^{local}<111$ and
$-151<f_{NL}^{equil}<253$.

\section{Stability Analysis}\label{sect-stab}
For model (\ref{action-Rin}) in the large curvature region, due to
the $1/R$ suppression, the correction term has negligible effects in
the early universe if $\mu$ is small \cite{Carroll:2003wy}. But
recent investigations \cite{Song:2006ej,Sawicki:2007tf} showed that
such correction terms may introduce instabilities, hence their
effects are not negligible even in the high redshift epoch. So it is
important to study the stability problem\footnote{We are grateful to
the referee for putting this problem to our attention.} in our model
(\ref{action-gRin}). Since the inflaton field is evolving, and there
is a signature change in $g(\varphi)$ around the end of inflation,
we should study this problem during inflation and after reheating
respectively.

To analyze the stability, there is an indicator $B$ given by formula
(17) in \cite{Song:2006ej} and formula (2) in \cite{Sawicki:2007tf}.
According to the results of \cite{Song:2006ej,Sawicki:2007tf}, the
instability resides in the branch of models with $B<0$. We find for
our model of the form (\ref{action-gRin}), the indicator
\begin{equation}\label{indicator}
B=\frac{2g}{M_p^2R^2-g}\frac{d\ln R}{d\ln a}\left(\frac{d\ln H}{d\ln a}\right)^{-1}.
\end{equation}
During inflation, for $g$ taking the form (\ref{gV}) and
$M_p^2V_{,\varphi\varphi}\ll V$, we get $B\simeq2>0$, so the model
is stable. This can also be inferred simply from the fact that $g>0$
during inflation.

As the inflaton rolled down the potential and decayed long after the
inflation, for the quadratic and quartic potentials, we have $V\ll
M_p^2V_{,\varphi\varphi}$ and then
\begin{equation}
B=-\frac{2V_{,\varphi\varphi}^2}{3R^2+V_{,\varphi\varphi}^2}\frac{d\ln R}{d\ln a}\left(\frac{d\ln H}{d\ln a}\right)^{-1}\leq0.
\end{equation}
The inequality is saturated if and only if $V_{,\varphi\varphi}=0$.
This cannot happen when $V=\frac{1}{2}m^2\varphi^2$ because it
always gives $V_{,\varphi\varphi}=m^2>0$. As a result, it does not
have a proper matter-dominated phase. This is another problem of the
model with a quadratic potential, as we have promised at the end of
section \ref{sect-quartic}.

But the story is a little different for the case
$V=\lambda\varphi^4$, in which the unstable branch with $B<0$ can be
avoided if the inflaton decayed to the minimum of its potential
$\varphi=0$ during the reheating era. Therefore, the stability
condition for the quartic potential model puts a constraint on
reheating: the reheating process should to efficient enough to
guarantee a complete decay of inflaton $\varphi$. Such an efficient
decay can be realized most easily by the instant preheating
mechanism \cite{Felder:1998vq}. After the complete decay of
$\varphi$ during reheating, the $f(\varphi,R)$ model of the form
(\ref{action-Rin4}) is reduced to the Einstein gravity without any
harmful instability in subsequent epochs.

The stability condition makes our models less interesting. Were the
stability condition ignored, one might drive the acceleration of
late universe with the residual non-vanishing inflaton, and thus
unify the two phases of accelerated expansion of the universe with a
single non-minimally coupled scalar field. After imposing the
stability condition, there is no room for such a natural
unification.

Of course, inspired by the so-called  mCDTT model in
\cite{Sawicki:2007tf}, one may replace $g(\varphi)$ with
$g(\varphi)+\mu^4M_p^2$ in (\ref{action-gRin}), where $\mu\ll M_p$
is a constant independent of $\varphi$. Then the newly added term
will play a role after inflation, exactly recover so-called mCDTT
model, which can avoid the instability problem. But, since the
constant $\mu$ is very small, this term plays no role during
inflation. Moreover, as was advocated in \cite{Amendola:2006kh}, all
$f(R)$ modified gravity during the matter phase is grossly
inconsistent with cosmological observations. So it seems that the
only choice for us is to recover Einstein gravity after reheating.

Therefore, although we started with the $f(R)$ model and the
accelerated expansion of late universe, due to various difficulties
put forward in
\cite{Amendola:2006kh,Amendola:2006eh,Song:2006ej,Sawicki:2007tf},
it turns out that our model has nothing to do with the $f(R)$ model
at the late time. In other words, the survived $f(\varphi,R)$
inflation model should reduce to the Einstein gravity after
reheating.

In contrast with \cite{Amendola:2006kh,Amendola:2006eh}, different
viewpoints are held by the authors of
\cite{Nojiri:2003ft,Nojiri:2006gh,Capozziello:2006dj}.\footnote{We
thank S. Odintsov and S. Nojiri for bringing
\cite{Nojiri:2003ft,Nojiri:2006gh,Capozziello:2006dj} to our
attention, and thank C. Corda for informing us about
\cite{Corda:2007hi,Corda:2009re}.} Interested readers may refine the
analysis above by taking
\cite{Nojiri:2003ft,Nojiri:2006gh,Capozziello:2006dj} into
consideration.

According to \cite{Corda:2007hi,Corda:2009re}, the interferometric
detection of gravitational waves can provide a definitive test for
general relativity. In other words, the interferometric detection of
gravitational waves will be a strong endorsement for the modified
gravity theories or, alternatively, will rule out them. So it would
be also necessary to further inspect the $f(\varphi,R)$ models from
this angle of view in the future.

\section{Comments and Conclusion}\label{sect-concl}
As we have stressed, our analysis throughout this paper is not more
than a semi-quantitative estimation. Before concluding, we would
like to remark on several weaknesses and the resulted uncertainties
in the above calculations. We can classify them into three
categories: the decoupled approximation inside the horizon, the
linear evolution approximation outside the horizon and the slow-roll
approximation.

First, as revealed by equations (\ref{evol-u}), the curvature
perturbation and the entropy perturbation are coupled inside the
horizon. But when writing down the analytical solution
(\ref{sol-u}), we have neglected the coupling terms. As a
subsequence, the correlation functions
$\langle\mathcal{R}_{\ast}\mathcal{S}_{\ast}\rangle$,
$\langle\mathcal{R}_{\ast}\mathcal{R}_{\ast}\mathcal{S}_{\ast}\rangle$
and
$\langle\mathcal{R}_{\ast}\mathcal{S}_{\ast}\mathcal{S}_{\ast}\rangle$
vanish only because we have neglected the coupling between
$\mathcal{R}_{\ast}$ and $\mathcal{S}_{\ast}$ inside the horizon.
All of the power spectra and three point functions at the
horizon-crossing should receive a correction from the coupling
effects. The correction is controlled by coupling coefficients
$\alpha$ and $\beta$ in evolution equations (\ref{evol-u}). This
is also a general problem for analytical solution of multi-field
inflation models. For a more accurate treatment to this problem in
two-field inflation, please refer to \cite{Byrnes:2006fr}.

Second, in deriving the transfer relation (\ref{spectra
transition}), we have neglected the nonlinear effects. As
mentioned in section \ref{sect-Gauss}, there are two sources of
nonlinear effects outside the horizon: the $\ddot{S}_k$ term
neglected in equation \eqref{trans-eqn}; the time dependence of
$\mu_{\mathcal{S}}$ and $\mu_{\mathcal{R}}$. Again, this is also a
general problem for analytical solution of multi-field inflation
models.

Third, there is an additional source of uncertainty for the model
studied in section \ref{sect-Rinverse}, where we have deliberately
kept the $V_{,\varphi\varphi}$ term in the Lagrangian. This is
necessary to avoid the divergence of power spectrum at the leading
order, but it brings some inconsistency for our slow-roll
approximation. This is clear from equations (\ref{slroll-Rin2}) and
(\ref{slroll-Rin4}), in which the slow-roll parameters are not of
the same order. In principle, this problem should be solved by doing
the calculations at the sub-leading order in a consistent way. But
the background dynamics will be rather messy, neither analytical nor
numerical method can give it a hand.

Although the analytic results obtained in this paper are not
accurate, it is still meaningful to take them for rough estimate
before painstaking calculation. There are some lessons we can learn
from it. For the $1/R$-corrected inflation, the evolution of entropy
perturbation can dramatically depress the tensor-to-scalar ratio and
enhance the magnitude of non-Gaussianity. Specifically, if we take
the rough estimation seriously, then the single-field inflation can
be rescued by the $1/R$ correction, otherwise it would have been
excluded by observational data.

The preliminary investigation in \cite{Ji:2009yw} and here raises
more questions than answers about generalized $f(\varphi,R)$ gravity
theories. First, we lack a first principle to write down the exact
form of $f(\varphi,R)$ when higher or lower order corrections are
considered. Second, all of the calculation makes sense only
semi-quantitatively, so a more accurate treatment is in demand. The
formalism we developed is applicable to other cosmological stages
and scenarios. Especially, it would be interesting to find a unified
model similar to \cite{Liddle:2008bm}, but with richer phenomena.
Third, it is possible that the entropy perturbation in
$f(\varphi,R)$ inflation can seed a tiny quadruple moment of
neutrinos, which deserves a detailed analysis. Fourth, according to
our rough estimate, the non-Gaussianity is large and positive in
some models. This is observationally interesting and should be
studied carefully in the future.

\acknowledgments{This work is supported by the China Postdoctoral
Science Foundation. We are grateful to acknowledge Xingang Chen,
Qing-Guo Huang and Yi Liao for helpful discussions. TW thanks the
hospitality of the Maryland Center for Fundamental Physics,
University of Maryland when this project was finished. The original
graph in Figure \ref{ns-vs-r} is downloaded from NASA website. We
acknowledge the use of the Legacy Archive for Microwave Background
Data Analysis (LAMBDA). Support for LAMBDA is provided by the NASA
Office of Space Science.}

\appendix

\section{Three-Point Correlations of the Local Form}\label{app-3pt}
Before engaging ourselves in calculation, we notice that the
definition of $\mathcal{R}_{k}$ and $\mathcal{S}_{k}$ in the text is
\begin{equation}\label{Fourier-def1}
\mathcal{R}(\mathbf{x},t)=\int\frac{d^3\mathbf{k}}{(2\pi)^{\frac{3}{2}}}e^{i\mathbf{k}\cdot\mathbf{x}}\mathcal{R}_{\mathbf{k}}(t),~~~~\mathcal{S}(\mathbf{x},t)=\int\frac{d^3\mathbf{k}}{(2\pi)^{\frac{3}{2}}}e^{i\mathbf{k}\cdot\mathbf{x}}\mathcal{S}_{\mathbf{k}}(t).
\end{equation}
But in calculating non-Gaussianity, usually a different
normalization is followed,
\begin{equation}\label{Fourier-def2}
\mathcal{R}(\mathbf{x},t)=\int\frac{d^3\mathbf{k}}{(2\pi)^3}e^{i\mathbf{k}\cdot\mathbf{x}}\tilde{\mathcal{R}}_{\mathbf{k}}(t),~~~~\mathcal{S}(\mathbf{x},t)=\int\frac{d^3\mathbf{k}}{(2\pi)^3}e^{i\mathbf{k}\cdot\mathbf{x}}\tilde{\mathcal{S}}_{\mathbf{k}}(t).
\end{equation}
In terms of $\tilde{\mathcal{R}}_{k}$ and
$\tilde{\mathcal{S}}_{k}$, we have the following relations between
two-point correlations and power spectra
\cite{Komatsu:2001rj,Komatsu:2002db}:
\begin{eqnarray}
\nonumber &&\langle\tilde{\mathcal{R}}_{\mathbf{k}1}\tilde{\mathcal{R}}_{\mathbf{k}2}\rangle=\frac{2\pi^2}{k^3}\tilde{\mathcal{P}}_{\mathcal{R}}(k)\delta(\mathbf{k}_1-\mathbf{k}_2)=\frac{(2\pi)^5}{2k^3}\mathcal{P}_{\mathcal{R}}(k)\delta(\mathbf{k}_1-\mathbf{k}_2),\\
\nonumber &&\langle\tilde{\mathcal{S}}_{\mathbf{k}1}\tilde{\mathcal{S}}_{\mathbf{k}2}\rangle=\frac{2\pi^2}{k^3}\tilde{\mathcal{P}}_{\mathcal{S}}(k)\delta(\mathbf{k}_1-\mathbf{k}_2)=\frac{(2\pi)^5}{2k^3}\mathcal{P}_{\mathcal{S}}(k)\delta(\mathbf{k}_1-\mathbf{k}_2),\\
\nonumber &&\langle\tilde{\mathcal{R}}_{\mathbf{k}1}\tilde{\mathcal{S}}_{\mathbf{k}2}\rangle=\frac{2\pi^2}{k^3}\tilde{\mathcal{C}}(k)\delta(\mathbf{k}_1-\mathbf{k}_2)=\frac{(2\pi)^5}{2k^3}\mathcal{C}(k)\delta(\mathbf{k}_1-\mathbf{k}_2),\\
\nonumber &&\langle\tilde{\mathcal{R}}(\mathbf{x}_1,t)\tilde{\mathcal{R}}(\mathbf{x}_2,t)\rangle=\int\frac{d^3\mathbf{k}}{4\pi k^3}\mathcal{P}_{\mathcal{R}}(k)e^{i\mathbf{k}\cdot(\mathbf{x}_1-\mathbf{x}_2)},\\
\nonumber &&\langle\tilde{\mathcal{S}}(\mathbf{x}_1,t)\tilde{\mathcal{S}}(\mathbf{x}_2,t)\rangle=\int\frac{d^3\mathbf{k}}{4\pi k^3}\mathcal{P}_{\mathcal{S}}(k)e^{i\mathbf{k}\cdot(\mathbf{x}_1-\mathbf{x}_2)},\\
&&\langle\tilde{\mathcal{R}}(\mathbf{x}_1,t)\tilde{\mathcal{S}}(\mathbf{x}_2,t)\rangle=\int\frac{d^3\mathbf{k}}{4\pi k^3}\mathcal{P}_{\mathcal{C}}(k)e^{i\mathbf{k}\cdot(\mathbf{x}_1-\mathbf{x}_2)}.
\end{eqnarray}

In accordance with the WMAP convention
\cite{Komatsu:2008hk,Komatsu:2001rj}, we parameterize the
nonlinearities of curvature and entropy perturbations as
\begin{eqnarray}
\nonumber \mathcal{R}(\mathbf{x},t)&=&\mathcal{R}_L-\frac{3}{5}f_{NL}^{\mathcal{R}}\left(\mathcal{R}_L^2-\langle\mathcal{R}_L^2\rangle\right),\\
\mathcal{S}(\mathbf{x},t)&=&\mathcal{S}_L-\frac{3}{5}f_{NL}^{\mathcal{S}}\left(\mathcal{S}_L^2-\langle\mathcal{S}_L^2\rangle\right).
\end{eqnarray}
Here $\mathcal{R}_L$ and $\mathcal{S}_L$ are linear Gaussian parts
of the perturbations. If we take nonlinear parameters
$f_{NL}^{\mathcal{R}}$ and $f_{NL}^{\mathcal{S}}$ as constants, then
this is a local form non-Gaussianity, which can be written in the
Fourier space as
\begin{eqnarray}
\nonumber \tilde{\mathcal{R}}(\mathbf{k})&=&\tilde{\mathcal{R}}_L(\mathbf{k})-\frac{3}{5}f_{NL}^{\mathcal{R}}\left[\int\frac{d^3\mathbf{p}}{(2\pi)^3}\tilde{\mathcal{R}}_L(\mathbf{k}-\mathbf{p})\tilde{\mathcal{R}}_L(\mathbf{p})-\int d^3\mathbf{x}e^{-i\mathbf{k}\cdot\mathbf{x}}\langle\tilde{\mathcal{R}}_L^2(\mathbf{x})\rangle\right],\\
\tilde{\mathcal{S}}(\mathbf{k})&=&\tilde{\mathcal{S}}_L(\mathbf{k})-\frac{3}{5}f_{NL}^{\mathcal{S}}\left[\int\frac{d^3\mathbf{p}}{(2\pi)^3}\tilde{\mathcal{S}}_L(\mathbf{k}-\mathbf{p})\tilde{\mathcal{S}}_L(\mathbf{p})-\int
d^3\mathbf{x}e^{-i\mathbf{k}\cdot\mathbf{x}}\langle\tilde{\mathcal{S}}_L^2(\mathbf{x})\rangle\right].
\end{eqnarray}
$\langle\mathcal{R}_L^2\rangle$ and $\langle\mathcal{S}_L^2\rangle$
are counter terms to ensure
$\langle\mathcal{R}(\mathbf{x},t)\rangle=\langle\mathcal{S}(\mathbf{x},t)\rangle=0$.

Using the above relations, it is straightforward to prove equation
(\ref{consist-rel}) and
\begin{eqnarray}
\nonumber &&\langle\tilde{\mathcal{R}}(\mathbf{k}_1)\tilde{\mathcal{R}}(\mathbf{k}_2)\tilde{\mathcal{S}}(\mathbf{k}_3)\rangle\\
\nonumber &=&-\frac{3}{5}f_{NL}^{\mathcal{R}}\int\frac{d^3\mathbf{p}}{(2\pi)^3}\langle\tilde{\mathcal{R}}_L(\mathbf{k}_1-\mathbf{p})\tilde{\mathcal{R}}_L(\mathbf{p})\tilde{\mathcal{R}}_L(\mathbf{k}_2)\tilde{\mathcal{S}}_L(\mathbf{k}_3)\rangle\\
\nonumber &&-\frac{3}{5}f_{NL}^{\mathcal{R}}\int\frac{d^3\mathbf{p}}{(2\pi)^3}\langle\tilde{\mathcal{R}}_L(\mathbf{k}_1)\tilde{\mathcal{R}}_L(\mathbf{k}_2-\mathbf{p})\tilde{\mathcal{R}}_L(\mathbf{p})\tilde{\mathcal{S}}_L(\mathbf{k}_3)\rangle\\
\nonumber &&-\frac{3}{5}f_{NL}^{\mathcal{S}}\int\frac{d^3\mathbf{p}}{(2\pi)^3}\langle\tilde{\mathcal{R}}_L(\mathbf{k}_1)\tilde{\mathcal{R}}_L(\mathbf{k}_2)\tilde{\mathcal{S}}_L(\mathbf{k}_3-\mathbf{p})\tilde{\mathcal{S}}_L(\mathbf{p})\rangle\\
\nonumber &&+\mathrm{divergent~counter~terms}\\
\nonumber &=&-\frac{6}{5}f_{NL}^{\mathcal{R}}\int d\mathbf{x}_1d\mathbf{x}_2d\mathbf{x}_3e^{-i\mathbf{k}_1\cdot\mathbf{x}_1-i\mathbf{k}_2\cdot\mathbf{x}_2-i\mathbf{k}_3\cdot\mathbf{x}_3}\langle\mathcal{R}_L(\mathbf{x}_1)\mathcal{R}_L(\mathbf{x}_2)\rangle\langle\mathcal{R}_L(\mathbf{x}_1)\mathcal{S}_L(\mathbf{x}_3)\rangle\\
\nonumber &&-\frac{6}{5}f_{NL}^{\mathcal{R}}\int d\mathbf{x}_1d\mathbf{x}_2d\mathbf{x}_3e^{-i\mathbf{k}_1\cdot\mathbf{x}_1-i\mathbf{k}_2\cdot\mathbf{x}_2-i\mathbf{k}_3\cdot\mathbf{x}_3}\langle\mathcal{R}_L(\mathbf{x}_1)\mathcal{R}_L(\mathbf{x}_2)\rangle\langle\mathcal{R}_L(\mathbf{x}_2)\mathcal{S}_L(\mathbf{x}_3)\rangle\\
\nonumber &&-\frac{6}{5}f_{NL}^{\mathcal{S}}\int d\mathbf{x}_1d\mathbf{x}_2d\mathbf{x}_3e^{-i\mathbf{k}_1\cdot\mathbf{x}_1-i\mathbf{k}_2\cdot\mathbf{x}_2-i\mathbf{k}_3\cdot\mathbf{x}_3}\langle\mathcal{R}_L(\mathbf{x}_1)\mathcal{S}_L(\mathbf{x}_3)\rangle\langle\mathcal{R}_L(\mathbf{x}_2)\mathcal{S}_L(\mathbf{x}_3)\rangle\\
\nonumber &=&-\frac{3}{10}(2\pi)^7f_{NL}^{\mathcal{R}}\frac{\mathcal{P}_{\mathcal{R}}(k_2)}{k_2^3}\frac{\mathcal{P}_{\mathcal{C}}(k_3)}{k_3^3}\delta^{(3)}(\mathbf{k}_1+\mathbf{k}_2+\mathbf{k}_3)\\
\nonumber &&-\frac{3}{10}(2\pi)^7f_{NL}^{\mathcal{R}}\frac{\mathcal{P}_{\mathcal{R}}(k_1)}{k_1^3}\frac{\mathcal{P}_{\mathcal{C}}(k_3)}{k_3^3}\delta^{(3)}(\mathbf{k}_1+\mathbf{k}_2+\mathbf{k}_3)\\
&&-\frac{3}{10}(2\pi)^7f_{NL}^{\mathcal{S}}\frac{\mathcal{P}_{\mathcal{C}}(k_1)}{k_1^3}\frac{\mathcal{P}_{\mathcal{C}}(k_2)}{k_2^3}\delta^{(3)}(\mathbf{k}_1+\mathbf{k}_2+\mathbf{k}_3).
\end{eqnarray}
By exchanging $\mathcal{R}\leftrightarrow\mathcal{S}$, one directly
writes down
\begin{eqnarray}
\nonumber &&\langle\tilde{\mathcal{S}}(\mathbf{k}_1)\tilde{\mathcal{S}}(\mathbf{k}_2)\tilde{\mathcal{R}}(\mathbf{k}_3)\rangle\\
\nonumber &=&-\frac{3}{10}(2\pi)^7f_{NL}^{\mathcal{S}}\frac{\mathcal{P}_{\mathcal{S}}(k_2)}{k_2^3}\frac{\mathcal{P}_{\mathcal{C}}(k_3)}{k_3^3}\delta^{(3)}(\mathbf{k}_1+\mathbf{k}_2+\mathbf{k}_3)\\
\nonumber &&-\frac{3}{10}(2\pi)^7f_{NL}^{\mathcal{S}}\frac{\mathcal{P}_{\mathcal{S}}(k_1)}{k_1^3}\frac{\mathcal{P}_{\mathcal{C}}(k_3)}{k_3^3}\delta^{(3)}(\mathbf{k}_1+\mathbf{k}_2+\mathbf{k}_3)\\
&&-\frac{3}{10}(2\pi)^7f_{NL}^{\mathcal{R}}\frac{\mathcal{P}_{\mathcal{C}}(k_1)}{k_1^3}\frac{\mathcal{P}_{\mathcal{C}}(k_2)}{k_2^3}\delta^{(3)}(\mathbf{k}_1+\mathbf{k}_2+\mathbf{k}_3).
\end{eqnarray}

The above derivation is valid when $f_{NL}^{\mathcal{R}}$ and
$f_{NL}^{\mathcal{S}}$ are constants. This is the case for the
local shape non-Gaussianity. We hope the results can be
generalized to other shapes as if $f_{NL}^{\mathcal{R}}$ and
$f_{NL}^{\mathcal{S}}$ are $\mathbf{k}$-dependent. But this
conjecture is to be proved or disproved by a more careful
investigation in the future.


\begin{thebibliography}{99}
\bibitem{Turyshev:2008ur}
  S.~G.~Turyshev,
  Usp.\ Fiz.\ Nauk {\bf 179}, 3 (2009)
  [Phys.\ Usp.\  {\bf 52}, 1 (2009)]
  [arXiv:0809.3730 [gr-qc]].

\bibitem{Riess:1998cb}
  A.~G.~Riess {\it et al.}  [Supernova Search Team Collaboration],
  Astron.\ J.\  {\bf 116}, 1009 (1998)
  [arXiv:astro-ph/9805201].

\bibitem{Perlmutter:1998np}
  S.~Perlmutter {\it et al.}  [Supernova Cosmology Project Collaboration],
  Astrophys.\ J.\  {\bf 517}, 565 (1999)
  [arXiv:astro-ph/9812133].

\bibitem{Komatsu:2008hk}
  E.~Komatsu {\it et al.}  [WMAP Collaboration],
  Astrophys.\ J.\ Suppl.\  {\bf 180}, 330 (2009)
  [arXiv:0803.0547 [astro-ph]].

\bibitem{Dvali:2000hr}
  G.~R.~Dvali, G.~Gabadadze and M.~Porrati,
  Phys.\ Lett.\  B {\bf 485}, 208 (2000)
  [arXiv:hep-th/0005016].

\bibitem{Carroll:2003wy}
  S.~M.~Carroll, V.~Duvvuri, M.~Trodden and M.~S.~Turner,
  Phys.\ Rev.\  D {\bf 70}, 043528 (2004)
  [arXiv:astro-ph/0306438].

\bibitem{Li:2004rb}
  M.~Li,
  Phys.\ Lett.\  B {\bf 603}, 1 (2004)
  [arXiv:hep-th/0403127].

\bibitem{Nojiri:2006ri}
  S.~Nojiri and S.~D.~Odintsov,
  eConf {\bf C0602061}, 06 (2006)
  [Int.\ J.\ Geom.\ Meth.\ Mod.\ Phys.\  {\bf 4}, 115 (2007)]
  [arXiv:hep-th/0601213].

\bibitem{Sotiriou:2008rp}
  T.~P.~Sotiriou and V.~Faraoni,
  arXiv:0805.1726 [gr-qc].

\bibitem{Ji:2009yw}
  X.~d.~Ji and T.~Wang,
  Phys.\ Rev.\  D {\bf 79}, 103525 (2009)
  [arXiv:0903.0379 [hep-th]].

\bibitem{Amendola:2006kh}
  L.~Amendola, D.~Polarski and S.~Tsujikawa,
  Phys.\ Rev.\ Lett.\  {\bf 98}, 131302 (2007)
  [arXiv:astro-ph/0603703].

\bibitem{Amendola:2006eh}
  L.~Amendola, D.~Polarski and S.~Tsujikawa,
  Int.\ J.\ Mod.\ Phys.\  D {\bf 16}, 1555 (2007)
  [arXiv:astro-ph/0605384].

\bibitem{Song:2006ej}
  Y.~S.~Song, W.~Hu and I.~Sawicki,
  Phys.\ Rev.\  D {\bf 75}, 044004 (2007)
  [arXiv:astro-ph/0610532].

\bibitem{Sawicki:2007tf}
  I.~Sawicki and W.~Hu,
  Phys.\ Rev.\  D {\bf 75}, 127502 (2007)
  [arXiv:astro-ph/0702278].

\bibitem{Nesseris:2008mq}
  S.~Nesseris,
  arXiv:0811.4292 [astro-ph].

\bibitem{Sadeghi:2009qr}
  J.~Sadeghi, M.~R.~Setare and A.~Banijamali,
  arXiv:0903.4073 [hep-th].

\bibitem{Bertolami:2009cd}
  O.~Bertolami and M.~C.~Sequeira,
  arXiv:0903.4540 [gr-qc].

\bibitem{Ito:2009nk}
  Y.~Ito and S.~Nojiri,
  arXiv:0904.0367 [hep-th].

\bibitem{Liddle:2008bm}
  A.~R.~Liddle, C.~Pahud and L.~A.~Urena-Lopez,
  Phys.\ Rev.\  D {\bf 77}, 121301 (2008)
  [arXiv:0804.0869 [astro-ph]].

\bibitem{Hwang:1990re}
  J.~C.~Hwang,
  Class.\ Quant.\ Grav.\  {\bf 7}, 1613 (1990).

\bibitem{Hwang:1996np}
  J.~C.~Hwang,
  Class.\ Quant.\ Grav.\  {\bf 14}, 1981 (1997)
  [arXiv:gr-qc/9605024].

\bibitem{Hwang:1996bc}
  J.~C.~Hwang,
  Class.\ Quant.\ Grav.\  {\bf 14}, 3327 (1997)
  [arXiv:gr-qc/9607059].

\bibitem{Hwang:1997uc}
  J.~C.~Hwang,
  Class.\ Quant.\ Grav.\  {\bf 15}, 1401 (1998)
  [arXiv:gr-qc/9710061].

\bibitem{Hwang:2005hb}
  J.~C.~Hwang and H.~Noh,
  Phys.\ Rev.\  D {\bf 71}, 063536 (2005)
  [arXiv:gr-qc/0412126].

\bibitem{Chen:2006wn}
  B.~Chen, M.~Li, T.~Wang and Y.~Wang,
  Mod.\ Phys.\ Lett.\  A {\bf 22}, 1987 (2007)
  [arXiv:astro-ph/0610514].

\bibitem{Ramirez:2009zs}
  E.~Ramirez and D.~J.~Schwarz,
  arXiv:0903.3543 [astro-ph.CO].

\bibitem{Teyssandier:1983zz}
  P.~Teyssandier and Ph.~Tourrenc,
  J.\ Math.\ Phys.\  {\bf 24}, 2793 (1983).

\bibitem{Maeda:1988ab}
  K.~I.~Maeda,
  Phys.\ Rev.\  D {\bf 39}, 3159 (1989).

\bibitem{Wands:1993uu}
  D.~Wands,
  Class.\ Quant.\ Grav.\  {\bf 11}, 269 (1994)
  [arXiv:gr-qc/9307034].

\bibitem{Matsuda:2009np}
  T.~Matsuda,
  arXiv:0906.0643 [hep-th].

\bibitem{Wands:2000dp}
  D.~Wands, K.~A.~Malik, D.~H.~Lyth and A.~R.~Liddle,
  Phys.\ Rev.\  D {\bf 62}, 043527 (2000)
  [arXiv:astro-ph/0003278].

\bibitem{Wands:2002bn}
  D.~Wands, N.~Bartolo, S.~Matarrese and A.~Riotto,
  Phys.\ Rev.\  D {\bf 66}, 043520 (2002)
  [arXiv:astro-ph/0205253].

\bibitem{Langlois:1999dw}
  D.~Langlois,
  Phys.\ Rev.\  D {\bf 59}, 123512 (1999)
  [arXiv:astro-ph/9906080].

\bibitem{Komatsu:2001rj}
  E.~Komatsu and D.~N.~Spergel,
  Phys.\ Rev.\  D {\bf 63}, 063002 (2001)
  [arXiv:astro-ph/0005036].

\bibitem{Chen:2006nt}
  X.~Chen, M.~x.~Huang, S.~Kachru and G.~Shiu,
  JCAP {\bf 0701}, 002 (2007)
  [arXiv:hep-th/0605045].

\bibitem{Maldacena:2002vr}
  J.~M.~Maldacena,
  JHEP {\bf 0305}, 013 (2003)
  [arXiv:astro-ph/0210603].

\bibitem{Li:2008gg}
  M.~Li and Y.~Wang,
  JCAP {\bf 0809}, 018 (2008)
  [arXiv:0807.3058 [hep-th]].

\bibitem{Komatsu:2002db}
  E.~Komatsu,
  arXiv:astro-ph/0206039.

\bibitem{Li:2008qc}
  M.~Li, T.~Wang and Y.~Wang,
  JCAP {\bf 0803}, 028 (2008)
  [arXiv:0801.0040 [astro-ph]].

\bibitem{Chen:2008wn}
  X.~Chen, R.~Easther and E.~A.~Lim,
  JCAP {\bf 0804}, 010 (2008)
  [arXiv:0801.3295 [astro-ph]].

\bibitem{Huang:2008ze}
  Q.~G.~Huang,
  Phys.\ Lett.\  B {\bf 669}, 260 (2008)
  [arXiv:0801.0467 [hep-th]].

\bibitem{Li:2008qv}
  S.~W.~Li and W.~Xue,
  arXiv:0804.0574 [astro-ph].

\bibitem{Gao:2008dt}
  X.~Gao,
  JCAP {\bf 0806}, 029 (2008)
  [arXiv:0804.1055 [astro-ph]].

\bibitem{Li:2008jn}
  M.~Li, C.~Lin, T.~Wang and Y.~Wang,
  arXiv:0805.1299 [astro-ph].

\bibitem{Li:2008fma}
  S.~Li, Y.~F.~Cai and Y.~S.~Piao,
  Phys.\ Lett.\  B {\bf 671}, 423 (2009)
  [arXiv:0806.2363 [hep-ph]].

\bibitem{Xue:2008mk}
  W.~Xue and B.~Chen,
  arXiv:0806.4109 [hep-th].

\bibitem{Huang:2008qf}
  Q.~G.~Huang,
  Phys.\ Rev.\  D {\bf 78}, 043515 (2008)
  [arXiv:0807.0050 [hep-th]].

\bibitem{Huang:2008rj}
  Q.~G.~Huang,
  JCAP {\bf 0809}, 017 (2008)
  [arXiv:0807.1567 [hep-th]].

\bibitem{Li:2008tw}
  M.~Li and C.~Lin,
  arXiv:0807.4352 [astro-ph].

\bibitem{Huang:2008bg}
  Q.~G.~Huang and Y.~Wang,
  JCAP {\bf 0809}, 025 (2008)
  [arXiv:0808.1168 [hep-th]].

\bibitem{Huang:2008zj}
  Q.~G.~Huang,
  JCAP {\bf 0811}, 005 (2008)
  [arXiv:0808.1793 [hep-th]].

\bibitem{Ling:2008xd}
  Y.~Ling and J.~P.~Wu,
  arXiv:0809.3398 [hep-th].

\bibitem{Moroi:2008nn}
  T.~Moroi and T.~Takahashi,
  Phys.\ Lett.\  B {\bf 671}, 339 (2009)
  [arXiv:0810.0189 [hep-ph]].

\bibitem{Tzirakis:2008qy}
  K.~Tzirakis and W.~H.~Kinney,
  JCAP {\bf 0901}, 028 (2009)
  [arXiv:0810.0270 [astro-ph]].

\bibitem{Gao:2009gd}
  X.~Gao and B.~Hu,
  arXiv:0903.1920 [astro-ph.CO].

\bibitem{Lyth:2005du}
  D.~H.~Lyth and Y.~Rodriguez,
  Phys.\ Rev.\  D {\bf 71}, 123508 (2005)
  [arXiv:astro-ph/0502578].

\bibitem{Lyth:2005fi}
  D.~H.~Lyth and Y.~Rodriguez,
  Phys.\ Rev.\ Lett.\  {\bf 95}, 121302 (2005)
  [arXiv:astro-ph/0504045].

\bibitem{Zaballa:2006pv}
  I.~Zaballa, Y.~Rodriguez and D.~H.~Lyth,
  JCAP {\bf 0606}, 013 (2006)
  [arXiv:astro-ph/0603534].

\bibitem{Cogollo:2008bi}
  H.~R.~S.~Cogollo, Y.~Rodriguez and C.~A.~Valenzuela-Toledo,
  JCAP {\bf 0808}, 029 (2008)
  [arXiv:0806.1546 [astro-ph]].

\bibitem{Rodriguez:2008hy}
  Y.~Rodriguez and C.~A.~Valenzuela-Toledo,
  arXiv:0811.4092 [astro-ph].

\bibitem{Komatsu:2009kd}
  E.~Komatsu {\it et al.},
  arXiv:0902.4759 [astro-ph.CO].

\bibitem{Byrnes:2006fr}
  C.~T.~Byrnes and D.~Wands,
  Phys.\ Rev.\  D {\bf 74}, 043529 (2006)
  [arXiv:astro-ph/0605679].

\bibitem{Weinberg:2004kr}
  S.~Weinberg,
  Phys.\ Rev.\  D {\bf 70}, 043541 (2004)
  [arXiv:astro-ph/0401313].

\bibitem{Weinberg:2004kf}
  S.~Weinberg,
  Phys.\ Rev.\  D {\bf 70}, 083522 (2004)
  [arXiv:astro-ph/0405397].

\bibitem{Felder:1998vq}
  G.~N.~Felder, L.~Kofman and A.~D.~Linde,
  Phys.\ Rev.\  D {\bf 59}, 123523 (1999)
  [arXiv:hep-ph/9812289].

\bibitem{Nojiri:2003ft}
  S.~Nojiri and S.~D.~Odintsov,
  Phys.\ Rev.\  D {\bf 68}, 123512 (2003)
  [arXiv:hep-th/0307288].

\bibitem{Nojiri:2006gh}
  S.~Nojiri and S.~D.~Odintsov,
  Phys.\ Rev.\  D {\bf 74}, 086005 (2006)
  [arXiv:hep-th/0608008].

\bibitem{Capozziello:2006dj}
  S.~Capozziello, S.~Nojiri, S.~D.~Odintsov and A.~Troisi,
  Phys.\ Lett.\  B {\bf 639}, 135 (2006)
  [arXiv:astro-ph/0604431].

\bibitem{Corda:2007hi}
  C.~Corda,
  JCAP {\bf 0704}, 009 (2007)
  [arXiv:astro-ph/0703644].

\bibitem{Corda:2009re}
  C.~Corda,
  arXiv:0905.2502 [gr-qc].
\end{thebibliography}
\end{document}